\documentclass[aps,prd,onecolumn,showkeys,amssymb]{revtex4}
 
\usepackage{float}
\usepackage{bm}
\usepackage{amsfonts}
\usepackage{latexsym}
\usepackage[latin1]{inputenc}
\usepackage{amssymb}
\usepackage{graphicx}
\usepackage{amsmath}
\usepackage{hyperref}
\usepackage{color}

\begin{document}

\newcommand{\be}{\begin{equation}}
\newcommand{\ee}{\end{equation}}
 
\title{From neutron stars to quark stars in mimetic gravity}
 
\author{Artyom V. Astashenok$^{1}$ and Sergei D. Odintsov$^{2,3}$}
 
\affiliation{$^{1}$I. Kant Baltic Federal University, Institute of Physics and Technology, Nevskogo st. 14, 236041 Kaliningrad, Russia,\\
$^2$Instituci\`{o} Catalana de Recerca i Estudis Avan\c{c}ats (ICREA), Barcelona, Spain\\
$^3$Institut de Ciencies de l'Espai (IEEC-CSIC), Campus UAB,
Facultat de Ciencies, Torre C5-Par-2a pl, E-08193 Bellaterra,
Barcelona, Spain\\}
 
\begin{abstract}
 
Realistic models of neutron and quark stars in the framework of
mimetic gravity with  Lagrange multiplier constraint are
presented.
 We discuss the
effect of mimetic scalar  aiming to describe dark matter on
mass-radius relation and the moment of inertia for slowly rotating
relativistic stars. The mass-radius relation and
 moment of inertia  depend on the value of mimetic scalar in the center of star. This fact leads to
the ambiguity in the mass-radius relation for a given equation of
state. {Such ambiguity allows to explain some observational facts
better than  in standard General Relativity}. The case of two
mimetic potentials namely $V(\phi)\sim A\phi^{-2}$ and
$V(\phi)\sim Ae^{B\phi^{2}}$ is considered in detail. The relative
deviation of maximal moment of inertia is approximately twice
larger than the relative deviation of maximal stellar mass. We
also briefly discuss the mimetic $f(R)$ gravity.
 In the case of $f(R)=R+aR^2$ mimetic gravity it is expected that increase of maximal mass and maximal moment of
inertia due to mimetic scalar  becomes much stronger with bigger
parameter $a$. The contribution of  scalar field in mimetic
gravity can lead to possible existence of extreme neutron stars
with large masses.
 
\end{abstract}
 

 
\maketitle
 
\section{Introduction}
 
Number of modified gravity theories which may describe
accelerating universe has been intensively studied in the recent
years (for a review, see \cite{review}). In the framework of these
theories it is possible to obtain the accelerated expansion of
Universe \cite{Perlmutter, Riess1, Riess2} without using of
inflaton for inflation and/or scalars, fluids or cosmological
constant for dark energy. Some cosmological bounds also favour the
modified gravity \cite{review}. However, one can not discriminate
between the $\Lambda$CDM model or the modified gravity using only
cosmological bounds.
 
The study of alternative gravities on the astrophysical level,
e.g. using the relativistic stars, allows for an alternative  way
of discrimination between General Relativity (GR) from its
possible modifications \cite{Dimitri-rev}. The fundamental
question is the  existence of stable neutron stars in given $f(R)$
theory \cite{Kobayashi-Maeda, Upadhye-Hu}. Secondly, one needs to
compare the mass-radius relation, moment of inertia, quadrupole
moment, Love number and other relevant characteristics of stars in
GR and $f(R)$ gravity \cite{Doneva2014a}. Finally, it is
interesting to consider the possible emergence of new stellar
structures in modified gravities (stable stars with large central
densities or/and with large magnetic fields, (super)massive stars
etc.). The discovery of such structures will  constitute a
powerful signature for the Extended Gravity \cite{Laurentis,
Laurentis2}. The structure of compact stars in perturbative $f(R)$
gravity was investigated recently in refs.\cite{Arapoglu,
Alavirad, Astashenok-1}. In this approach the scalar curvature $R$
is defined by Einstein equations at zeroth order on the small
parameter, i.e. $R\sim T$, where $T$ is the trace of
energy-momentum tensor. Non-perturbative studies are also
available for non-rotating, slowly and fast rotating compact star
models \cite{Yazadjiev2014,Capo}.
 
In this paper we investigate relativistic stars in mimetic gravity
with scalar potential $V(\phi)$ (mimetic potential) and with
Lagrange multiplier $\beta(\phi)$. This theory was recently
proposed in ref.\cite{10} for eventual geometric description of
dark matter. The  paper is organized as  follows. In the next
section, we present the field equations for mimetic $f(R)$ gravity
with mimetic scalar potential. For spherically symmetric solutions
of these equations,  the modified Tolman--Oppenheimer--Volkoff
(TOV) equations are derived. The corresponding scalar-tensor
description of mimetic $f(R)$ theory is given in the third
section. Mimetic $f(R)$ theory is equivalent to a two
scalar-tensor
 gravity.
 In Section IV, the compact star models in frame of simple
mimetic gravity with $f(R)=R$ are investigated in detail. We
obtain the mass-radius and mass-moment of inertia relations for
neutron stars in the case of simple mimetic potentials. For
neutron stars we use a well-known equation of state proposed by
Douchin and Haensel (Sly4 EoS) \cite{DH2000}. We also investigate
the case of quark stars with simple EoS for deconfined quark
matter. Some conclusions are given in Summary.
 
\section{Modified TOV equations in mimetic $f(R)$ gravity with  scalar potential}
 
The main feature of mimetic approach to gravity is that the
conformal symmetry is a non-violated internal degree of freedom.
The physical metric $g_{\mu\nu}$ can be represented via so-called
auxiliary metric as \cite{10} 
\be \label{cond}
g_{\mu\nu}=-\hat{g}^{\rho\sigma}\partial_{\rho}\phi
\partial_{\sigma}\phi \hat{g}_{\mu\nu}.
 \ee
Here $\phi$ and $\hat{g}_{\mu\nu}$ are mimetic field and an
auxiliary metric tensor correspondingly. Then one takes variation
with respect to the auxiliary metric $\hat{g}_{\mu\nu}$ and to the
scalar field $\phi$ instead of the physical metric. This variation
gives the following condition on auxiliary scalar field $\phi$: 
\be {g}^{\rho\sigma}(\hat{g}_{\mu\nu},\phi)\partial_{\rho}\phi
\partial_{\sigma}\phi=-1. \ee
Such parametrization is invariant under Weyl transformations
$\hat{g}_{\mu\nu} = e^{\sigma(x)}g_{\mu\nu}$ and therefore the
auxiliary metric tensor doesn't appear in the action for the
gravitational field.
 
For spherically symmetric solution describing the star one needs
to consider a metric of the following form
\begin{equation}\label{metric}
    ds^2= -e^{2\psi}dt^2 +e^{2\lambda}dr^2 +r^2 d\Omega^2 \, .
\end{equation}
The metric functions $\psi$ and $\lambda$ depend only from radial
coordinate. In addition, we  assume that the auxiliary scalar
field depends only on the $r$. The action for mimetic $f(R)$
theory with a scalar potential $V(\phi)$ and a Lagrange multiplier
$\beta$ in the Jordan frame (in units where $G=c=1$) is given as
\cite{mpla}:
\begin{equation}\label{action}
S=\frac{1}{16\pi}\int d^4x
\sqrt{-g}\left[f(R(g_{\mu\nu})-V(\phi)+\beta(g^{\mu\nu}\partial_{\mu}\phi
\partial_{\nu}\phi+1)\right] + S_{{\rm matter}},
\end{equation}
where $g$ is determinant of the metric $g_{\mu\nu}$ and $S_{\rm
matter}$ is the action of the standard perfect fluid matter. Note
that above theory may consistently unify the early-time inflation
with late-time acceleration and geometric dark matter\cite{mpla}.
The accelerating cosmology of such mimetic gravity was
investigated recently in refs.\cite{mim}. It is also interesting
that mimetic scalar may eventually play the role of chameleon
\cite{justin}.

Varying the action with respect to  $g_{\mu\nu}$ gives us the
equation of motion for metric functions: 
\begin{eqnarray}\label{field}
f'(R)G_{\mu \nu }-\frac{1}{2}(f(R)-f'(R)R)g_{\mu \nu } &-&(\nabla
_{\mu }\nabla _{\nu }-g_{\mu \nu }\Box )f'(R) \nonumber \\
&=& 8 \pi T_{\mu \nu
}+g_{\mu\nu}\left(-V(\phi)+\beta(g^{\rho\sigma}\partial_{\rho}\phi
\partial_{\sigma}\phi+1)) -2\beta\partial_{\mu}\phi
\partial_{\nu}\phi \right).
\end{eqnarray}
Here $G_{\mu\nu}=R_{\mu\nu}-\frac{1}{2}Rg_{\mu\nu}$ is the
Einstein tensor, $f'(R)=df(R)/dR$ is the derivative of function
$f(R)$ with respect to the scalar curvature $R$ and $T_{\mu \nu }$
is the energy--momentum tensor. For perfect fluid we have
$T_{\mu\nu}=\mbox{diag}(e^{2\psi}\rho, e^{2\lambda}p, r^2p,
r^{2}\sin^{2}\theta p)$, where $\rho$ is the matter density and
$p$ is the pressure.
 
Assuming that $\phi=\phi(r)$ and varying the action with respect
to $\beta$ one can obtain the following constraint for the
auxiliary field: 
\be e^{-2\lambda}\left(\frac{d\phi}{dr}\right)^{2}+1=0. \ee 
One may consider substituting $\phi$ as follows
$\phi\rightarrow\phi^{*}=i\phi$. 
Therefore we have the following equation for $\phi^{*}$: 
\be\label{constr}
e^{-2\lambda}\left(\frac{d\phi^{*}}{dr}\right)^{2}=1 \ee 
which in the limit  $r\rightarrow\infty$ becomes
$\left(\frac{d\phi^{*} }{dr}\right)^{2}\rightarrow 1$.
 
The Tolmann-Oppenheimer-Volkov (TOV) equations for this theory of
gravity are nothing else that ``$tt$'' and ``$rr$'' components of
the field equations (\ref{field}):
 
\begin{eqnarray} \label{TOV1}
\frac{f'(R)}{r^2}\frac{d}{dr}\left [r\left(1-e^{-2\lambda
}\right)\right]&=&8\pi \rho+V(\phi)+
+\frac{1}{2}\left(f'(R)R-f(R)\right) \nonumber \\
&+&e^{-2\lambda}\left[\left(\frac{2}{r}-\frac{d\lambda}{dr}\right)\frac{d
f'(R)}{dr}+\frac{d^{2}f'(R)}{dr^2}\right]
\end{eqnarray}
 
 
\begin{eqnarray}
 \label{TOV2}
 \frac{f'(R)}{r}
\left[2e^{-2\lambda}\frac{d\psi}{dr}-\frac{1}{r}\left(1-e^{-2\lambda}\right)\right]
&=&8\pi
p-V(\phi)-\beta\left(e^{-2\lambda}\left(\frac{d\phi}{dr}\right)^2-1\right)+\nonumber \\
&+&\frac{1}{2}\left(f'(R)R-f(R)\right)+e^{-2\lambda}\left(\frac{2}{r}+\frac{d\psi}{dr}\right)\frac{df'(R)}{dr}
\end{eqnarray}
For the scalar  $\phi$ we have equation of motion 
\be\label{auxil}
2\triangledown^{\mu}(\beta\partial_{\mu}\phi)+\frac{dV(\phi)}{d\phi}=0.
\ee 
The hydrostatic  equilibrium conditionfollows from the
conservation law, $\nabla^\mu T_{\mu\nu}=0$ i.e. 
\begin{equation}\label{hydro}
    \frac{dp}{dr}=-(\rho
    +p)\frac{d\psi}{dr}.
\end{equation}
In f(R) gravity the scalar curvature is dynamical variable and an
equation for $R$ can be obtained using the trace of field equation
(\ref{field}). One gets
\begin{equation}\label{TOV3}
3\square f'(R)+f'(R)R-2f(R)=-{8\pi}(\rho-3p)-4V(\phi)+2\beta,
\end{equation}
where 
$$
e^{2\lambda}\square=\left(\frac{2}{r}+\frac{d\psi}{dr}-\frac{d\lambda}{dr}\right)\frac{d}{dr}+\frac{d^2}{dr^2}
\, .
$$
For $f(R)=R$ this equation is reduced to the form
\be \label{R0} R={8\pi}(\rho-3p)+4V(\phi)-2\beta \, . \ee 
Inside the star the equations (\ref{TOV1}), (\ref{TOV2}),
(\ref{hydro}), (\ref{TOV3}) can be solved numerically for a given
equation of state (EOS) $p=f({\rho})$ and boundary conditions
$\lambda(0)=0$, $R(0)=R_{c}$, $R'(0)=0$ and
${\rho}(0)={\rho}_{c}$.
 
Outside the star ($\rho=p=0$) the solution is defined by the Eqs.
(\ref{TOV1}), (\ref{TOV2}), (\ref{TOV3}), while on the surface of
star ($r=r_{s}$) the junction conditions should be satisfied:
$$\lambda_{in}(r_{s})=\lambda_{out}(r_{s}),\quad
R_{in}(r_{s})=R_{out}(r_{s}), \quad R'_{in}(r_{s})=R'_{out}(r_{s})
\, .
$$
 
The gravitational mass parameter $m(r)$ is defined as
\begin{equation}
\label{mass}
    e^{-2\lambda}=1-\frac{2m}{r}.
\end{equation}
 
Finally, the asymptotic flatness requirement gives the constraints
on scalar curvature and mass parameter:
$$\lim_{r\rightarrow\infty}R(r)=0,
\lim_{r\rightarrow\infty}m(r)=\mbox{const}.$$

\section{Scalar-tensor description for mimetic $f(R)$ gravity}
 
By analogy with convenient $F(R)$ one can  consider  mimetic
$f(R)$ theory in the Einstein frame. In this case the theory is
just a two scalar-tensor gravity. We start from the equivalent
Brans-Dicke  action:
 \be
 S_{g}=\frac{1}{16\pi}\int
d^{4} x \sqrt{-g}\left(\Phi R -
U(\Phi)-V(\phi)+\beta(\phi)(g^{\mu\nu}\partial_{\mu}\phi
\partial_{\nu}\phi+1)\right).
\ee 
Here $\Phi=df(R)/dR$ is gravitational scalar and
$U(\Phi)=Rf'(R)-f(R)$ is a potential. Using conformal
transformation $\tilde{g}_{\mu\nu}=\Phi g_{\mu\nu}$ one can write
the action in the Einstein frame as follows: 
 \be
 S_{g}=\frac{1}{16\pi}\int d^{4} x
\sqrt{-\tilde{g}}\left(\tilde{R}
-2\tilde{g}^{\mu\nu}\partial_{\mu}\varphi\partial_{\nu}\varphi-4\tilde{V}(\varphi)-V(\phi)e^{-4\varphi/\sqrt{3}}+\beta(\phi)
e^{-2\varphi/\sqrt{3}}(\tilde{g}^{\mu\nu}\partial_{\mu}\phi
\partial_{\nu}\phi+e^{-2\varphi/\sqrt{3}})\right),
\ee 
where $\varphi=\frac{\sqrt{3}}{2}\ln \Phi$ and the redefined
potential ${V}(\varphi)$ in Einstein frame  becomes
$\tilde{V}(\varphi)=\Phi^{-2}(\varphi)U(\Phi(\varphi))/4$.
 
The form of the spacetime metric can be chosen to  coincide
formally with the GR form (\ref{metric}): 
\be \label{metric2} d\tilde{s}^{2}=\Phi
ds^{2}=-e^{2\tilde{\psi}}dt^{2}+e^{2\tilde{\phi}}\tilde{dr}^{2}+\tilde{r}^2d\Omega^2.
\ee 
In Eq. (\ref{metric2}) $\tilde{r}^2=\Phi r^{2}$,
$e^{2\tilde{\psi}}=\Phi e^{2\psi}$ and from the equality $ \Phi
e^{2\lambda}dr^{2}=e^{2\tilde{\lambda}}d\tilde{r}^{2} $ follows
that
\begin{equation}
e^{-2\lambda}=e^{-2\tilde{\lambda}}\left(1-\tilde{r}\varphi'(\tilde{r})/\sqrt{3}\right)^{2}
\, .
\end{equation}
Therefore the mass parameter $m(r)$ can be obtained from
$\tilde{m}(\tilde{r})$ and gets the form: 
\be m(\tilde{r})=
\frac{\tilde{r}}{2}\left(1-\left(1-\frac{2\tilde{m}}{\tilde{r}}\right)\left(1-\tilde{r}\varphi'(\tilde{r})/\sqrt{3}\right)^{2}\right)e^{-\varphi/\sqrt{3}}
\ee 
The resulting equations for metric functions $\tilde{\lambda}$ and
$\tilde{\psi}$ coincide in fact with ordinary TOV equations for
mimetic General Relativity in which the energy density and
pressure of the scalar field $\varphi$ are included (the tildes
are omitted for simplicity): 
\begin{eqnarray}
 \label{TOV1-1}
 \frac{1}{r^2}\frac{d}{dr}\left
[r\left(1-e^{-2\lambda }\right)\right] &=&
e^{-4\varphi/\sqrt{3}}(8\pi\rho+V(\phi))+e^{-2\lambda}\left(\frac{d\varphi}{dr}\right)^{2}+\tilde{V}(\varphi)
\, ,
\\
 \label{TOV2-1}
 \frac{1}{r} \left[2e^{-2\lambda}\frac{d\psi}{dr}-\frac{1}{r}\left(1-e^{-2\lambda}\right)\right]
&=& e^{-4\varphi/\sqrt{3}}\left(8\pi p-\beta(\phi)
(e^{-2\lambda}e^{2\varphi/\sqrt{3}}\left(\frac{d\phi}{dr}\right)^{2}-1)\right)+e^{-2\lambda}\left(\frac{d\varphi}{dr}\right)^{2}-\tilde{V}(\varphi)
\, .
\end{eqnarray}
The hydrostatic equilibrium condition is written as 
\begin{equation}\label{hydro-1}
    \frac{dp}{dr}=-(\rho+p)\left(\frac{d\psi}{dr}-\frac{1}{\sqrt{3}}\frac{d\varphi}{dr}\right).
\end{equation}
The equation (\ref{auxil}) for the mimetic field becomes:
\be 2\triangledown^{\mu}(\beta
e^{-2\varphi/\sqrt{3}}\partial_{\mu}\phi)+V'(\phi)e^{-4\varphi/\sqrt{3}}=0.
\ee 
 Finally, the last equation of
motion for the scalar field is equivalent to Eq. (\ref{TOV3}) in
$f(R)$ theory: 
\be \label{TOV3-1} \square
\varphi+\frac{dV(\varphi)}{d\varphi}=-\frac{4\pi}{\sqrt{3}}
e^{-4\varphi/\sqrt{3}}(\rho-3p)-\frac{1}{\sqrt{3}}
e^{-4\varphi/\sqrt{3}}(2V(\phi)-\beta) . \ee 
The above formulation maybe used to study relativistic stars in
the Einstein frame. Unfortunately, due to number of issues (the
appearence of singular points where mathematical equivalence is
lost, the issue of boundary conditions in different frames, wider
space of physical variables(negative values of scalar)) the
physical equivalence with Jordan frame may not be explicit. To
avoid all above problems,  eventually it is better to work  with
$F(R)$ frame.

\section{Relativistic stars in mimetic General Relativity}
 
Let us study models of relativistic compact stars in mimetic
General Relativity. Even in this case some interesting deviations
from convenient GR can be observed.

\subsection{Neutron stars}
 
There are many equations of state for dense nuclear matter
 (see for example \cite{Camenzind,Potekhin}). One of the most commonly used is the
 SLy4 EOS. Here we use an analytic fitting of the EOS, see \cite{Camenzind}:
\begin{eqnarray}\label{SLY4}
\zeta&=&\frac{a_{1}+a_{2}\xi+a_{3}\xi^3}{1+a_{4}\xi}f(a_{5}(\xi-a_{6}))+(a_{7}+a_{8}\xi)f(a_{9}(a_{10}-\xi)) \\
&+&(a_{11}+a_{12}\xi)f(a_{13}(a_{14}-\xi))+(a_{15}+a_{16}\xi)f(a_{17}(a_{18}-\xi)),
\end{eqnarray}
where 
$$
\zeta=\log(P/\mbox{dyn} \mbox{cm}^{-2})\,, \qquad
\xi=\log(\rho/\mbox{g}\mbox{cm}^{-3})\,, \qquad
f(x)=\frac{1}{\exp(x)+1}\,
$$
and $a_{i}$ are some coefficients.
 
Let us consider the case of potential $V(\phi)=A\phi^{-2}$. Note
that $V(\phi^{*})=-A\phi^{*-2}=A^{*}\phi^{*-2}$ ($A^{*}=-A$) and
$V'(\phi)=2iA^{*}\phi^{*-3}$. From Eq. (\ref{auxil}) we obtain
equation for real field $\phi^{*}$: 
\be
2\triangledown^{\mu}(\beta\partial_{\mu}\phi^{*})+2A^{*}\phi^{*-3}=0.
\ee 
In the rest of the work we   omit the asterisk for simplicity. For
positive (negative) values of $\phi$ in the center of star one
needs to choose the positive (negative) root in Eq. (\ref{constr})
for $d\phi/dr$ (in the opposite case singularity occurs because
scalar field $\phi \rightarrow 0$). Without loss of generality one
may assume that $\phi(r)>0$.
 
The mass-radius (M-R) relation depends on the choice of the value
for the scalar mimetic field in center of star, $\phi(0)$. {Thus
the M-R curves are parametrized by $\phi(0)$, while we assume
$\beta(0)=0$. } For $A>0$ the mass of star for given radius
increases with decrease of $\phi(0)$. The results of calculations
are given in Table I. In Fig.~1 the mass-radius diagram and the
dependence of the stellar  mass on the central density are
presented for $A=0.005$ and three values of  $\phi(0)$. It is
noticeable that the mass-radius relation differs significantly
from the corresponding in GR for small masses i.e. $M<\sim
1.4M_\odot$. Also in the low-density regime ($\rho_c<4\times
10^{14}$ g/cm$^{3}$) the mimetic gravity models acquire
significantly larger masses than  GR ones due to scalar
contribution.

\begin{table}
\label{Table1}
\begin{centering}
\caption{Neutron star models (for SLy4 EoS) with
$V(\phi)=A\phi^{-2}$ for various values of $A$ and $\phi(0)$. For
comparison the GR maximal mass for this EoS is $2.05M_{\odot}$
($\rho_{c}=2.86\times 10^{15}$ g/cm$^{3}$, $R=$9.98 km). }
\begin{tabular}{|c|c|c|c||c|c|c|c||c|c|c|c|}
\hline
\multicolumn{4}{l}{$A=0.005$}   & \multicolumn{4}{l}{$A=0.01$}  & \multicolumn{4}{l}{$A=0.02$}  \\
\hline
$\phi(0)$ & $M_{max}$, &  $R$,  & $\rho_c$,  & $\phi(0)$ & $M_{max}$, &  $R$, & $\rho_c$,  &$\phi(0)$ & $M_{max}$, &  $R$,  & $\rho_c$,\\
  &  $M_{\odot}$ & km & 10$^{15}$ g/cm$^{3}$  & & $M_{\odot}$ & km & 10$^{15}$ g/cm$^{3}$  & & $M_{\odot}$ & km & 10$^{15}$ g/cm$^{3}$\\
  \hline
  0.10 & 2.08  & 10.09 & 2.75   & 0.1 & 2.11  & 10.20 & 2.45   &0.5 & 2.11  & 10.16 & 2.45  \\
  0.05 & 2.09 & 10.15 & 2.64   &0.05 & 2.13 & 10.41 & 2.26   &0.25 & 2.14 & 10.35 & 2.26  \\
  0.01 & 2.11  & 10.33 & 2.35   &  0.01 & 2.19  & 10.57 & 1.86   &0.05 & 2.24 & 10.87 & 1.65  \\
  \hline
\end{tabular}
 
\end{centering}
 
\end{table}
 

\begin{figure}
  \includegraphics[scale=0.7]{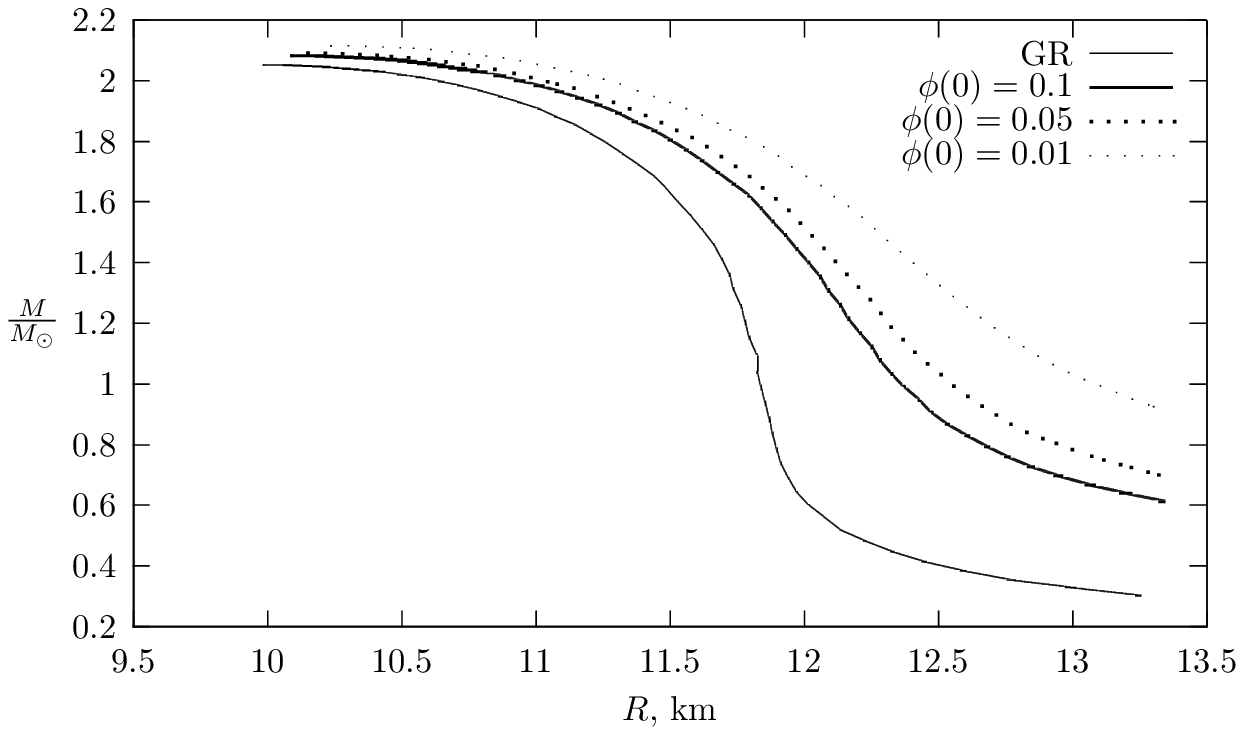} \includegraphics[scale=0.7]{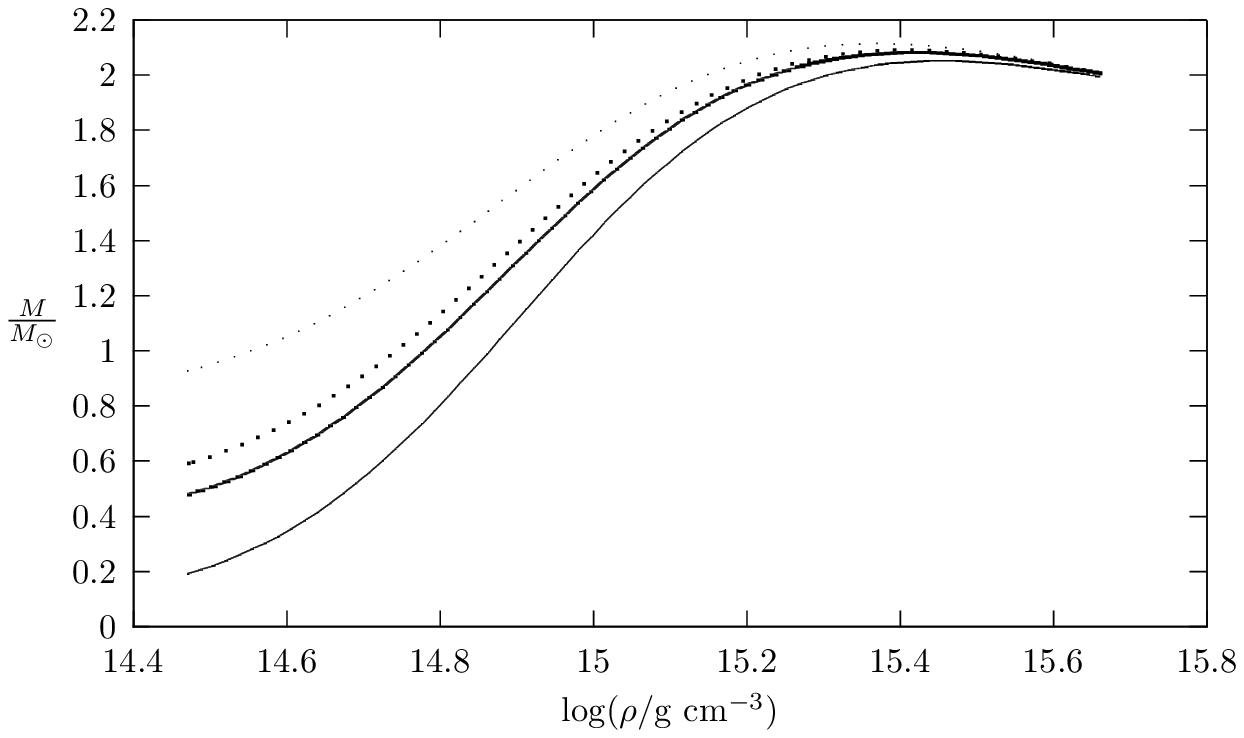}
\caption{The mass-radius diagram (left panel) and dependence of
neutron star mass on the central density (right panel) for neutron
stars in mimetic GR with $V(\phi)=A\phi^{-2}$ ($A=0.005$) in
comparison with GR by using a SLy4 equation of state for various
values of $\phi(0)$.}
\end{figure}
For the case of exponential potential $V(\phi)=Ae^{C \phi^{2}}$
the maximal mass is determined mainly by value of parameter $A$
and weakly depends from $\phi(0)$ (see Table II and Fig. 2).
 
\begin{table}
\label{Table1}
 
\begin{centering}
\caption{Neutron star models for SLy4 EoS with $V(\phi)=Ae^{C
\phi^{2}}$ for various values of $A$ and $\phi(0)$ ($C=-0.5$).}
\begin{tabular}{|c||c|c|c|c|c|c|}
\hline
\multicolumn{4}{l}{\hskip 1cm $A=0.02$}  & \multicolumn{3}{l}{\hskip 0.3cm$A=0.04$}  \\
\hline
$\phi(0)$ & $M_{max}$, &  $R$,  & $\rho_c$,  & $M_{max}$, &  $R$,  & $\rho_c$,\\
  &  $M_{\odot}$ & km & 10$^{15}$ g/cm$^{3}$ &    $M_{\odot}$ & km & 10$^{15}$ g/cm$^{3}$\\
  \hline
  0.10 & 2.08  & 10.12 & 2.64  & 2.12  & 10.32 & 2.35 \\
  0.05 & 2.09 & 10.10 & 2.64   & 2.12 & 10.32 & 2.35\\
  0.01 & 2.09  & 10.19 & 2.54 & 2.13  & 10.32 & 2.35 \\
  \hline
\end{tabular}
\end{centering}
 
\end{table}
 
\begin{figure}
  \includegraphics[scale=0.7]{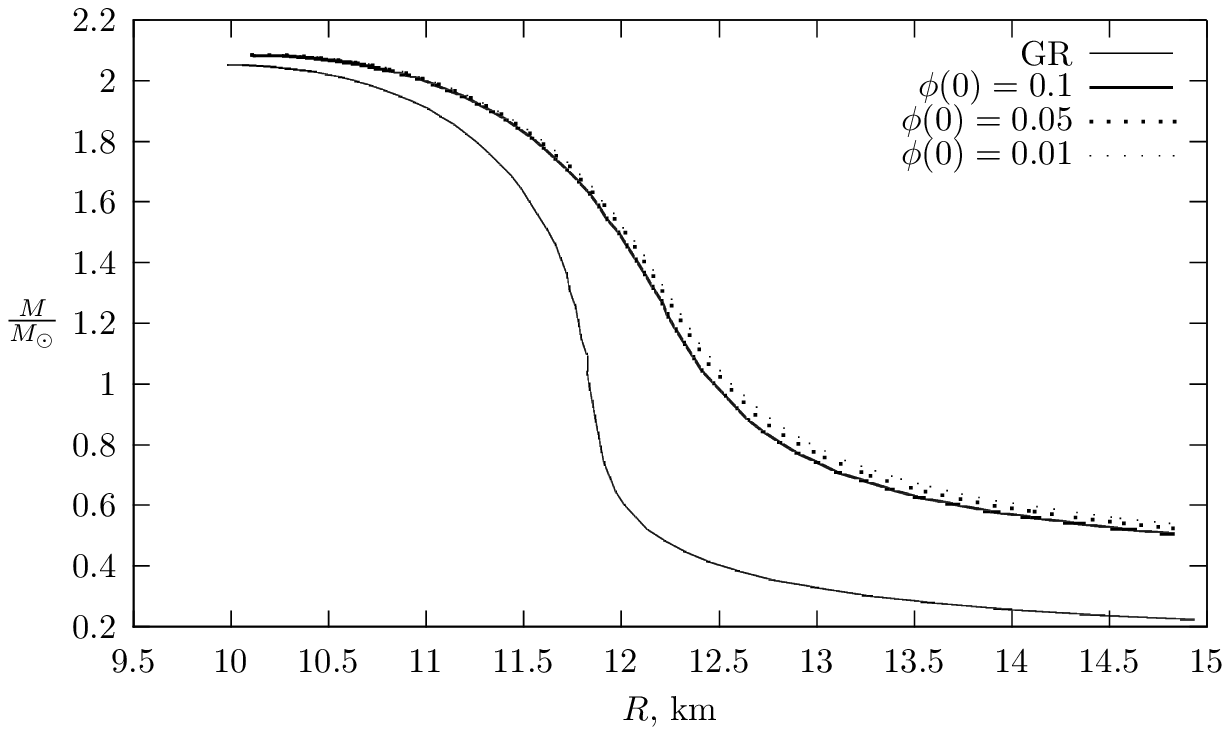} \includegraphics[scale=0.7]{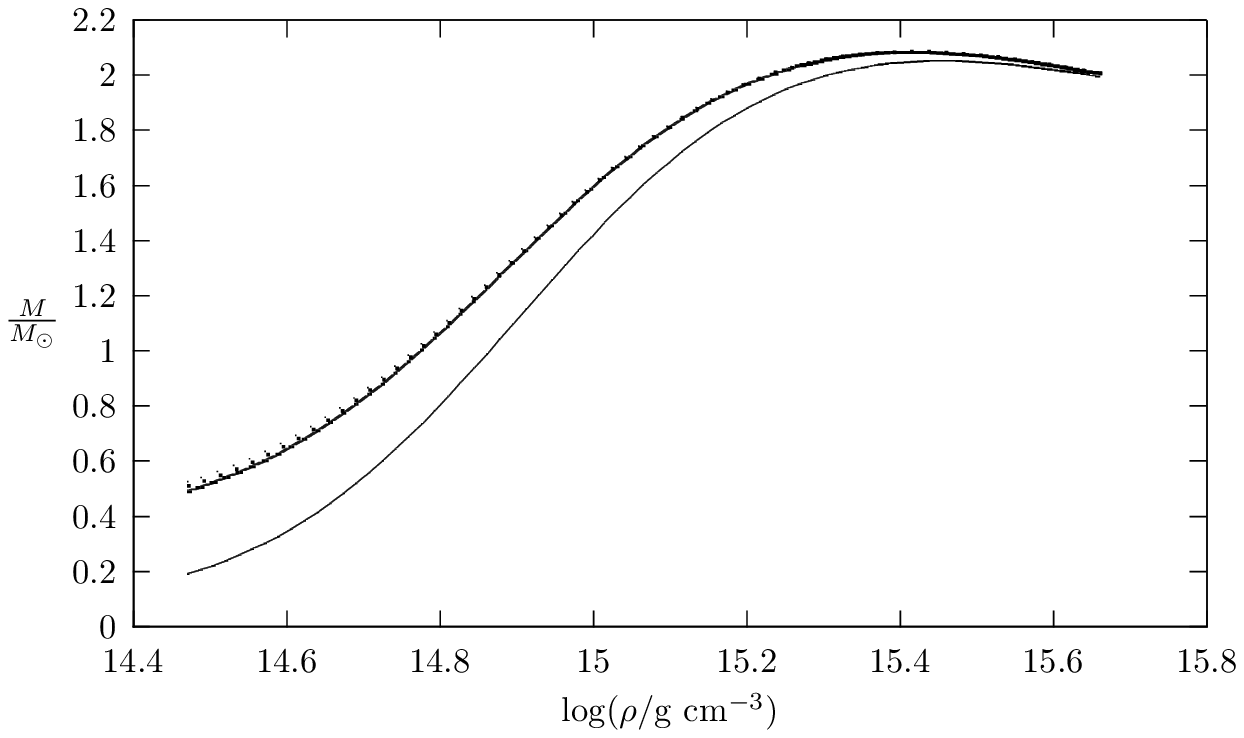}\\
\caption{The mass-radius diagram (left panel) and dependence of
neutron star mass on central density (right panel) for neutron
stars in mimetic GR with $V(\phi)=Ae^{C \phi^{2}}$ ($A=0.02$,
$C=0.5$) in comparison with GR by using the SLy4 equation of state
for various values of $\phi(0)$.}
\end{figure}
 
The presence of the scalar field affects the maximal  mass limit.
Still the effect is not of the same scale as it has been found for
the lower density neutron stars. The effect of the auxiliary field
can be more significant for stiffer equations of state (smaller
compactness) or/and in the presence of rotation.
 
The ambiguity of mass-radius relation is determined by the free
parameter $\phi(0)$ which can explain some inconsistencies in
$M-R$ relation from observations and theoretical considerations.
For example, from the study of longer X-ray bursters
\cite{Suleimanov, Hambaryan} it follows that neutron stars with
masses $M\sim 1-1.4 M_{\odot}$ have large radii $R>14.0$ km.
However, the analysis of both transiently accreting and bursting
sources \cite{Lattimer2012} suggests that the radius of a $1.4
M_\odot$ neutron star should not exceed the $12.9$ km. These
contradictions may indicate that the neutron star mass is
determined not only by the EOS of the dense matter but also by
other parameters. In mimetic gravity these free parameters are
present. Of course, there are many other EOS for which neutron
stars can have larger radius for the typical $1.4 M_\odot$ models.
However, our main point is that even EoS which is considered to be
not fully realistic due to discovery of large mass neutron mass,
may still be viable in modified gravity! The results presented
here provide hint that unique discrimination between General
Relativity and mimetic gravity can be made once we know in detail
the equation of state.

\subsection{Quark stars}
 
Here we consider the properties of quark stars \cite{Itoh, Witten}
in mimetic gravity. The hypothetical quark stars  consist of
deconfined light quarks ($u$, $d$ and $s$) and electrons. The
equation for quark matter that forms a colour supeconductor is
quite simple in frames of MIT bag model:
\be \label{QEOS}p=\alpha (\rho-4B). \ee
Here $B$ is the so called ``bag constant''. The value of parameter
$\alpha$ depends on the chosen mass of strange quark and varies
from $\alpha=1/3$ ($m_s=0$) to $\alpha=0.28$ ($m_{s}=250$ MeV).
The value of $B$ lies in interval $0.98<B<1.52$ in units of
$B_{0}=60$ MeV/fm$^{3}$ \cite{Sterg}. It is worth noticing that
the discovery of neutron stars with masses of the order of
$\approx 2 M_\odot$ \cite{Demorest2010,Antoniadis2013} imposes
severe constraints on the equation of state (EOS) of nuclear
matter and especially on the possibility of existence of quark
stars.

We consider two cases: 1) $\alpha=0.28$, $B=B_{0}$ and 2)
$\alpha=1/3$, $B=B_{0}$ assuming the potential
$V(\phi)=A\phi^{-2}$ . For quark stars the mass-radius relation is
significantly different from the  GR one for large masses (see
Table III and Fig. 3 for results). The deviation from General
Relativity is stronger in comparison with neutron stars (for same
$A$ and $\phi(0)$).
 
\begin{table}
\label{Table3}
\begin{centering}
\caption{Quark star models with $V(\phi)=A\phi^{-2}$ for various
$A$ and $\phi(0)$. For comparison in General Relativity maximal
mass for considered EoS is $1.77M_{\odot}$ ($\rho_{c}=2.26\times
10^{15}$ g/cm$^{3}$, $R=10.23$ km) for $\alpha=0.28$ and
$1.97M_{\odot}$ ($\rho_{c}=2.01\times 10^{15}$ g/cm$^{3}$,
$R=10.76$ km) for $\alpha=1/3$.}
\begin{tabular}{|c||c|c|c||c|c|c|}
\multicolumn{4}{c}{$\alpha=0.28$} & \multicolumn{3}{c}{$\alpha=1/3$} \\
\hline
\multicolumn{6}{l}{$A=0.005$}  \\
\hline
$\phi(0)$ & $M_{max}$, &  $R$,  & $\rho_c$,  & $M_{max}$, &  $R$,  & $\rho_c$, \\
  &  $M_{\odot}$  & km & 10$^{15}$ g/cm$^{3}$ &   $M_{\odot}$  & km & 10$^{15}$ g/cm$^{3}$ \\
  \hline
  0.1 & 1.84  & 10.42 & 1.93  & 2.04  & 10.90 & 1.79 \\
  0.05 & 1.86 & 10.50 & 1.79  & 2.06 & 10.99 & 1.65 \\
  0.01 & 1.91  & 10.66 & 1.47 & 2.10  & 11.07 & 1.47 \\
  \hline
  \multicolumn{6}{l}{$A=0.01$}  \\
    \hline
  0.1 & 1.91  & 10.62 & 1.65  & 2.11  & 11.01 & 1.53  \\
  0.05 & 1.96 & 10.78 & 1.41  & 2.15 & 11.23 & 1.36 \\
  0.01 & 2.07  & 11.08 & 0.99 & 2.26  & 11.51 & 0.99 \\
  \hline
\end{tabular}
 
\end{centering}
\end{table}
 
\begin{figure}
  \includegraphics[scale=0.7]{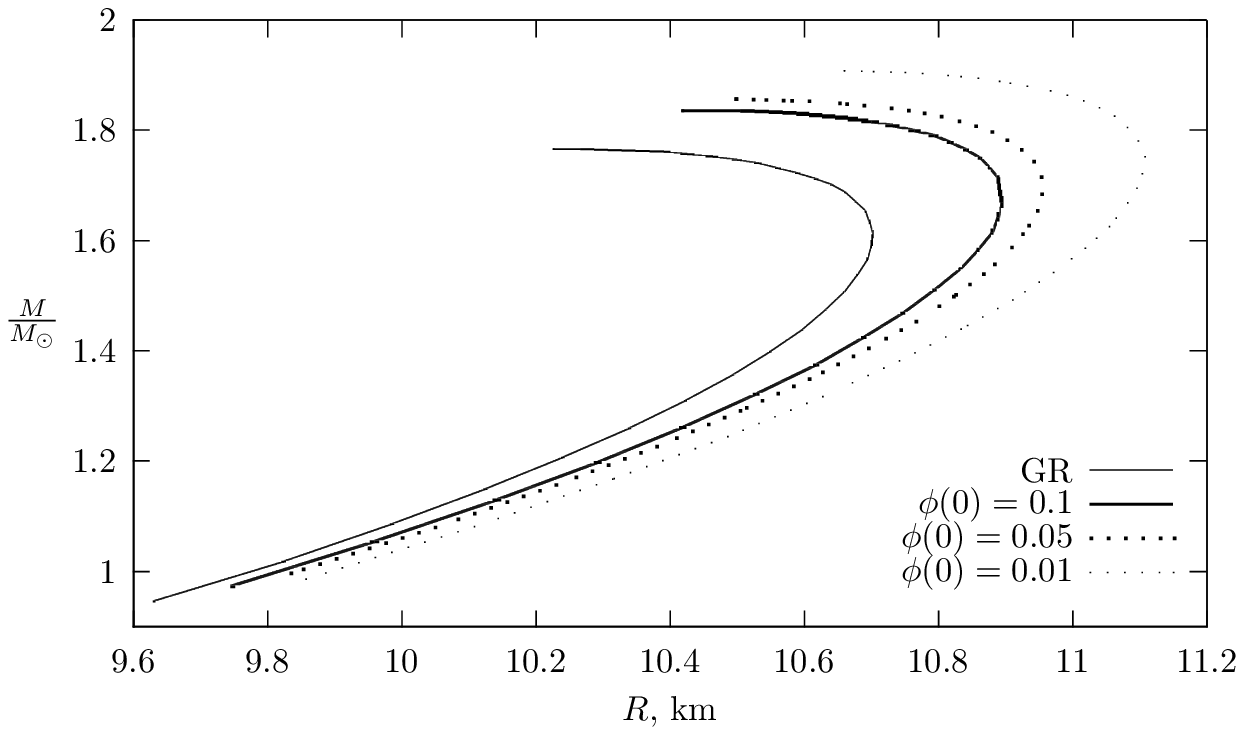} \includegraphics[scale=0.7]{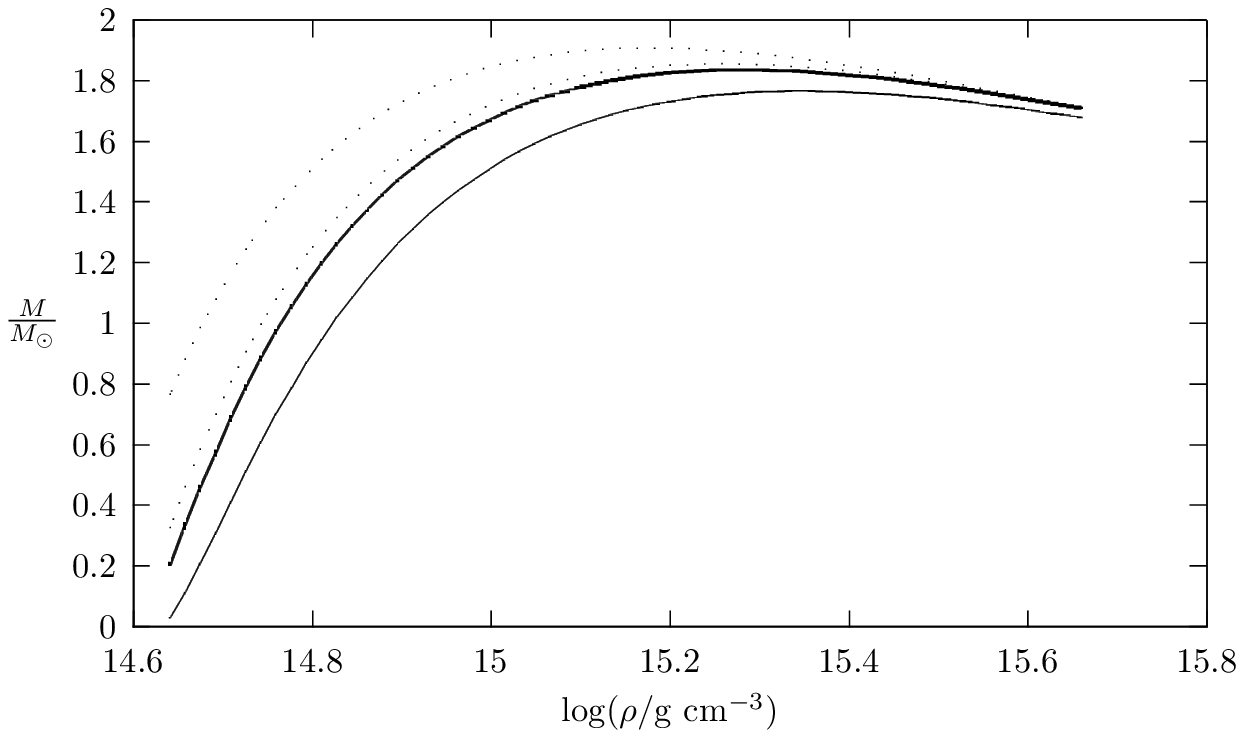}\\
  \includegraphics[scale=0.7]{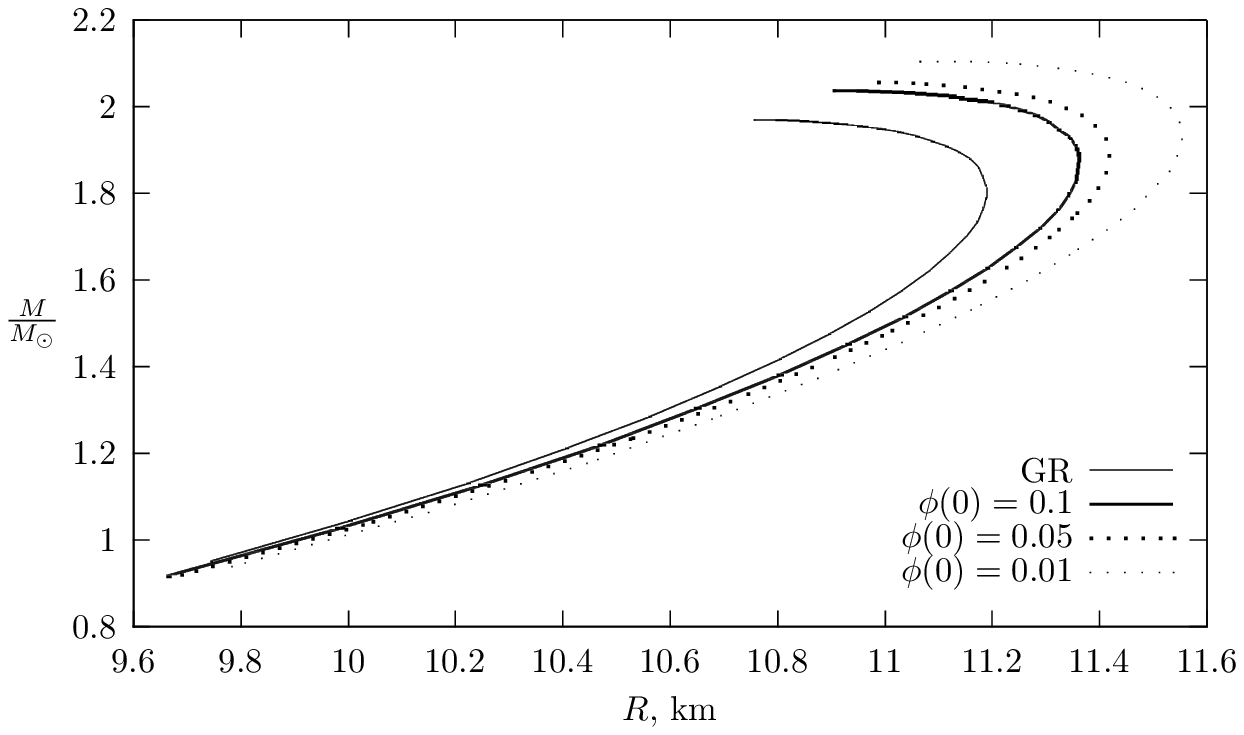} \includegraphics[scale=0.7]{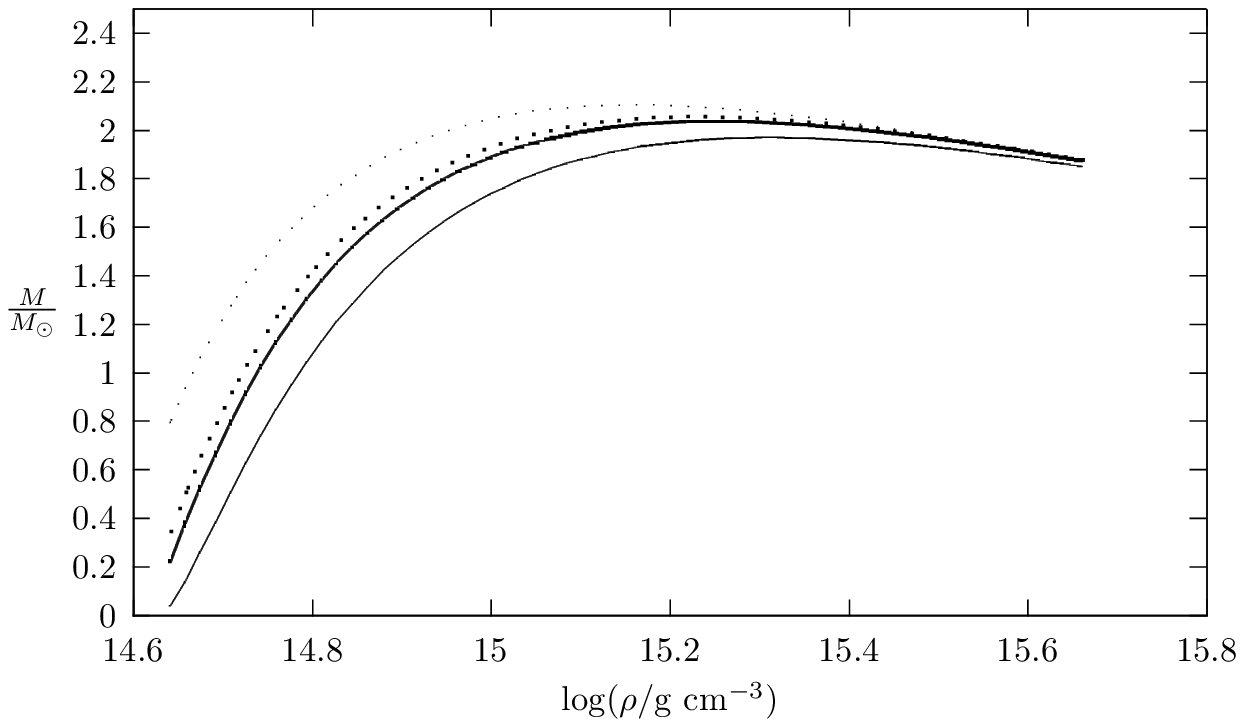}\\
\caption{The mass-radius diagram (left) and the dependence of mass
on the central density (right) for quark stars in mimetic GR with
$V(\phi)=A\phi^{-2}$ ($A=0.005$) in comparison with General
Relativity. Here we used
  a simple EoS (\ref{QEOS}) with $\alpha=0.28$, $B=B_{0}$ (upper panel)
  and $\alpha=1/3$,
  $B=B_{0}$ (lower panel)
   for various values
  of $\phi(0)$.}
\end{figure}
 
For $V(\phi)=Ae^{C \phi^{2}}$ the results are similar to above
considered for neutron stars. The deviation from General
Relativity is defined mainly by value of parameter $A$ (see Fig.
4).
\begin{figure}
  \includegraphics[scale=0.7]{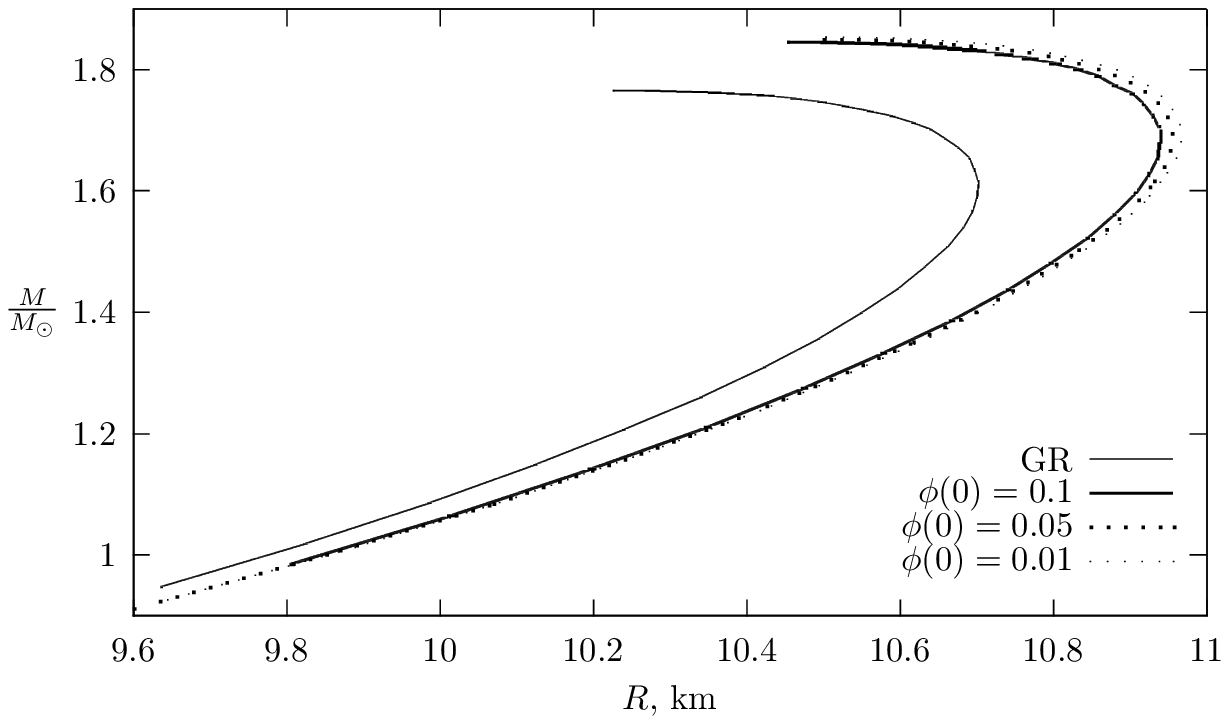} \includegraphics[scale=0.7]{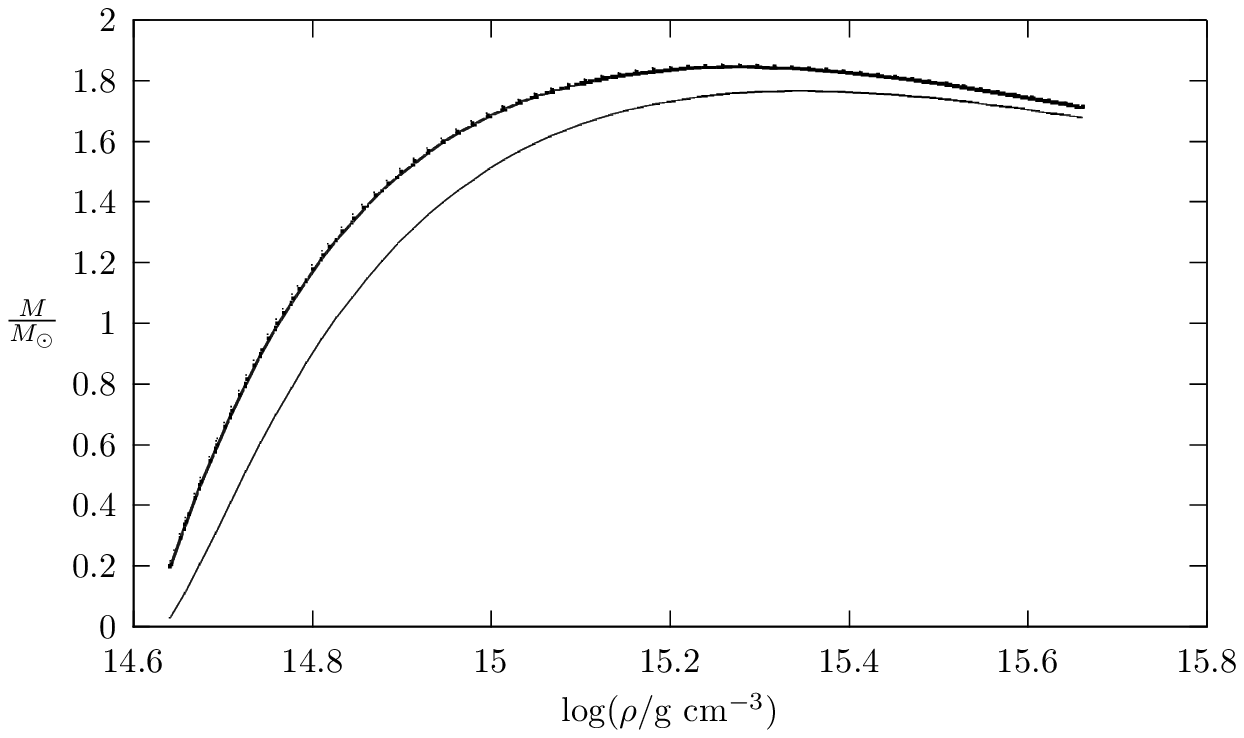}\\
  \includegraphics[scale=0.7]{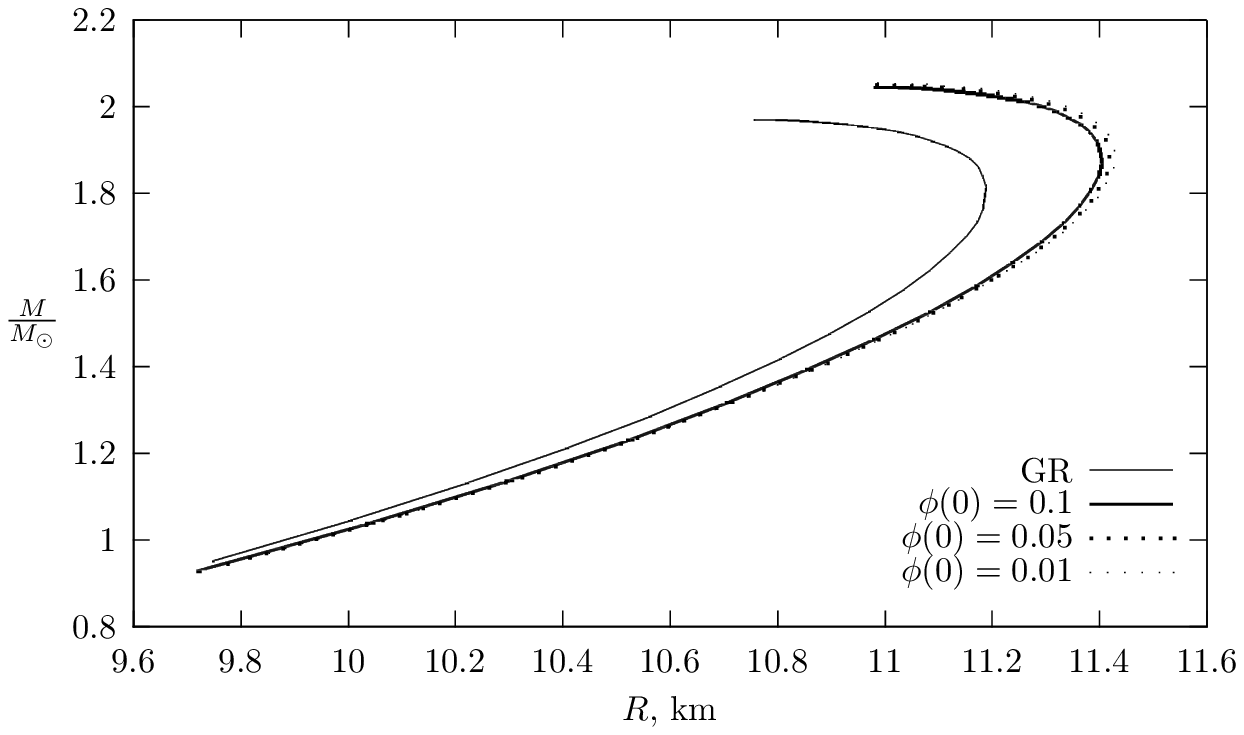} \includegraphics[scale=0.7]{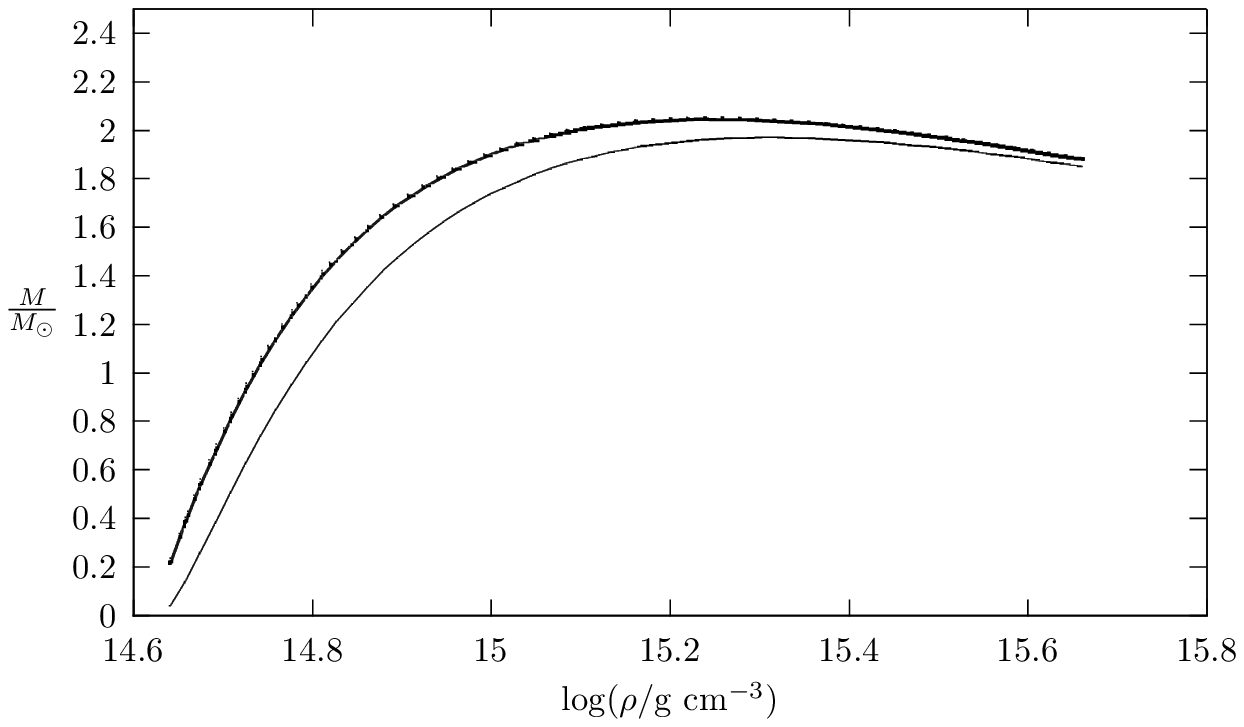}\\
  \caption{The mass-radius diagram (left) and dependence of mass on the central density (right) for quark stars in mimetic GR with $V(\phi)=Ae^{C \phi^{2}}$ ($A=0.02$,
$C=-0.5$) in comparison with General Relativity. Here we used  the
EOS (\ref{QEOS}) with $\alpha=0.28$, $B=B_{0}$ (upper panel) and
$\alpha=1/3$,
  $B=B_{0}$ (lower panel)  for various values of $\phi(0)$.
  The results seem to be insensitive to the central values of the $\phi(0)$ because the auxiliary field increases quickly and $V(\phi)\rightarrow 0$.
In fact the mass-radius relation depends solely on the parameter
$A$ and not on $\phi(0)$.}
\end{figure}
 
The quark star models under investigation have the maximum  mass
below the maximum known mass $\approx 2M_{\odot}$. However, in the
framework of mimetic gravity with a simple potential the maximal
mass can meet and exceed this limit set by observations.
 
\subsection{The inertial characteristics of compact stars in mimetic gravity.}
 
It is interesting to investigate slowly rotating compact stars in
mimetic gravity with potential $V(\phi)=A\phi^{-2}$.
 
The spacetime metric  with only first-order rotational terms with
respect to ${\Omega}=u^{\phi}/u^{t}$ has the form
\begin{eqnarray}
ds^2= - e^{2\psi(r)}dt^2 + e^{2\lambda(r)}dr^2 +
r^2\left(d\theta^2+\sin^{2}\theta\left(d\phi-(\Omega-\omega(r))dt\right)^{2}\right).
\end{eqnarray}

For slowly rotating stars the moment of inertia  is given by 
\be
I=-\frac{2}{3}\int_{0}^{R}r^{3}\frac{dj}{dr}\frac{\omega(r)}{\Omega},
\ee
where $R$ is the radius of non-rotating model and $
j(r)=e^{-\lambda(r)-\psi(r)}. $
 
In the scalar-tensor theory of gravity one needs to add an extra
equation for the function $\omega(r)$ to system (\ref{TOV1-1})--
(\ref{TOV3-1}). Taking only first-order terms on $\omega(r)$, the
field equations in the Einstein frame can be written as 
\begin{equation}\label{TOV4-1}
\frac{e^{\psi-\lambda}}{r^4} \partial_{r}\left[e^{-(\psi +
\lambda)} r^4 \partial_{r}{\omega} \right]  +
\frac{1}{r^2\sin^3\theta}
\partial_{\theta}\left[\sin^3\theta\partial_{\theta}\omega
\right]= 16\pi e^{-4\varphi/\sqrt{3}}(\rho + p)\omega.
\end{equation}
In asymptotically flat spacetime ${\omega}$ is a function of
radial coordinate only. Therefore Eq. (\ref{TOV4-1}) can be
rewritten as 
\begin{equation}\label{OR}
\frac{e^{\psi-\lambda}}{r^4} \frac{d}{dr}\left[e^{-(\psi+
\lambda)}r^4 \frac{d{\omega}(r)}{dr} \right] = 16\pi
e^{-4\varphi/\sqrt{3}}(\rho + p){\omega}(r).
\end{equation}
The function $\omega(r)$ should obey the following two boundary
conditions. 
Firstly, the condition of regularity at the $r=0$ which requires
\begin{equation}
\frac{d{\omega}(0)}{dr}= 0.
\end{equation}
Secondly,
\begin{equation}
\lim_{r\to \infty}{\omega}={\Omega}.
\end{equation}

Then the relation for the moment of inertia  can be written as:
\begin{equation}\label{Inertial}
I= \frac{8\pi}{3} \int_{0}^{r_s}e^{-4\varphi/\sqrt{3}}(\rho +
p)e^{\lambda - \psi} r^4 \left(\frac{\omega}{{\Omega}}\right) dr .
\end{equation}

For comparison, we calculate the relative deviation of the maximum
mass and the maximum moment of inertia in mimetic gravity and
General Relativity: 
$$\Delta M_{\rm max}=\frac{M_{\rm max}-M^{GR}_{\rm
max}}{M^{GR}_{\rm max}}, \quad \Delta I_{\rm max}=\frac{I_{\rm
max}-I^{GR}_{\rm max}}{I^{GR}_{\rm max}}.
$$
The results for neutron and quark stars are given in Tables IV and
V correspondingly. The deviations of maximal mass and moment of
inertia as function of $\phi_{0}=\phi(0)$ are given on Fig. 5. For
given $A$ there is minimal value of $\phi_{0}$ such that for
$\phi(0)<\phi_{0}$ stable stars cannot exist.
 
On Fig. 6 (left panel) the moment of inertia as function of mass
is plotted for $A=0.005$ for considered models of neutron and
quark stars. For given values of $A$ and $\phi(0)$ the increase of
maximal moment of inertia and maximal mass are much stronger than
for neutron stars.
 
We also examined the normalized metric function $\omega(r)/\Omega$
as a function of the radial coordinate.  On Fig. 6 this function
is depicted for two star configurations with $M=1.4M_\odot$ and
mass close to $M^{GR}_{\rm max}$ for each EoS. One can see that
deviations from General Relativity are larger in the vicinity of
star core, this is a feature noticed also in \cite{Staykov2014}
for slowly rotating neutron star models in $R^2$ gravity. The
contribution of auxiliary scalar field grows with decreasing
$\phi_{0}$ and therefore this leads to stronger deviations from
General Relativity.
 
\begin{figure}
a)\\
\includegraphics[scale=0.7]{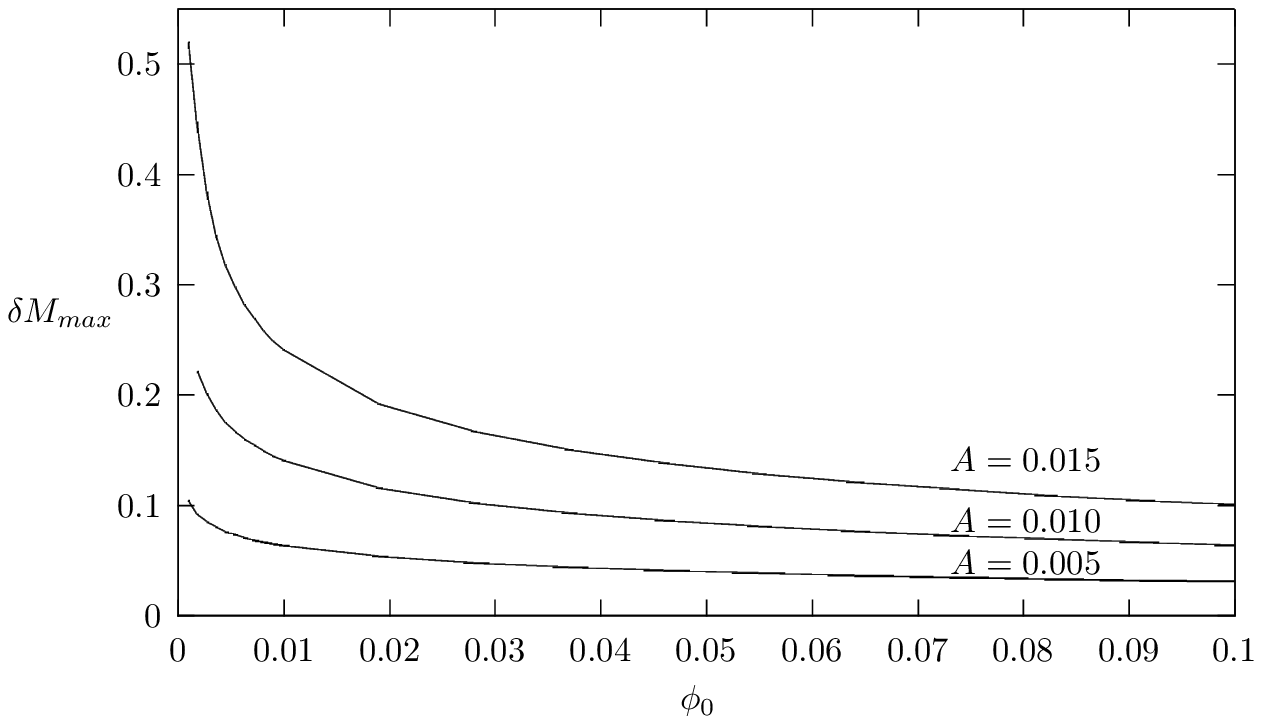} \includegraphics[scale=0.7]{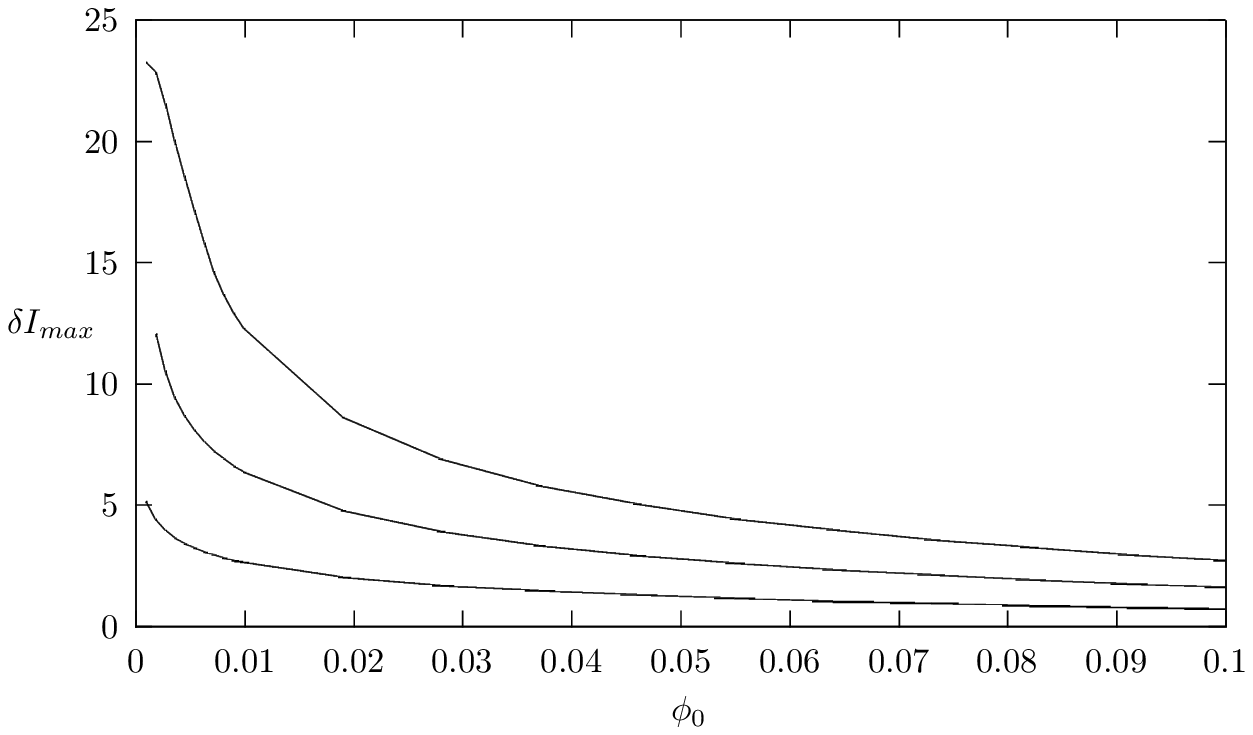}\\
b)\\
\includegraphics[scale=0.7]{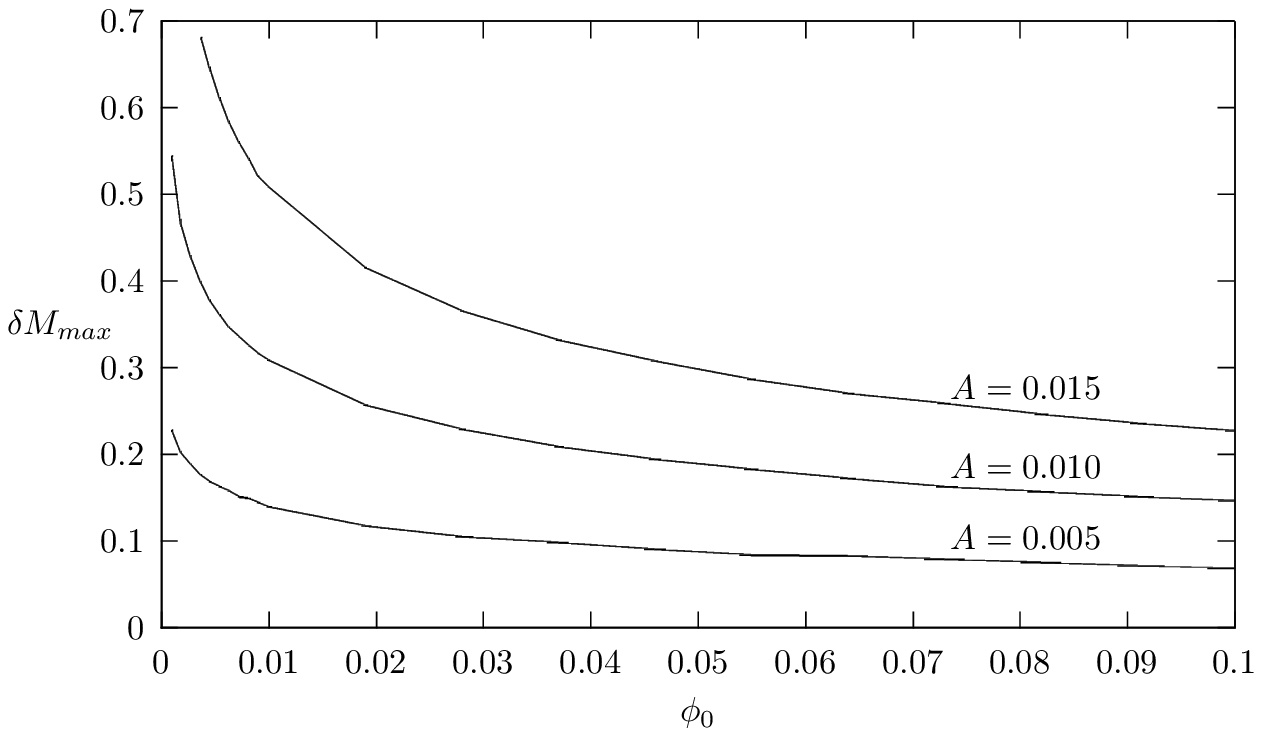} \includegraphics[scale=0.7]{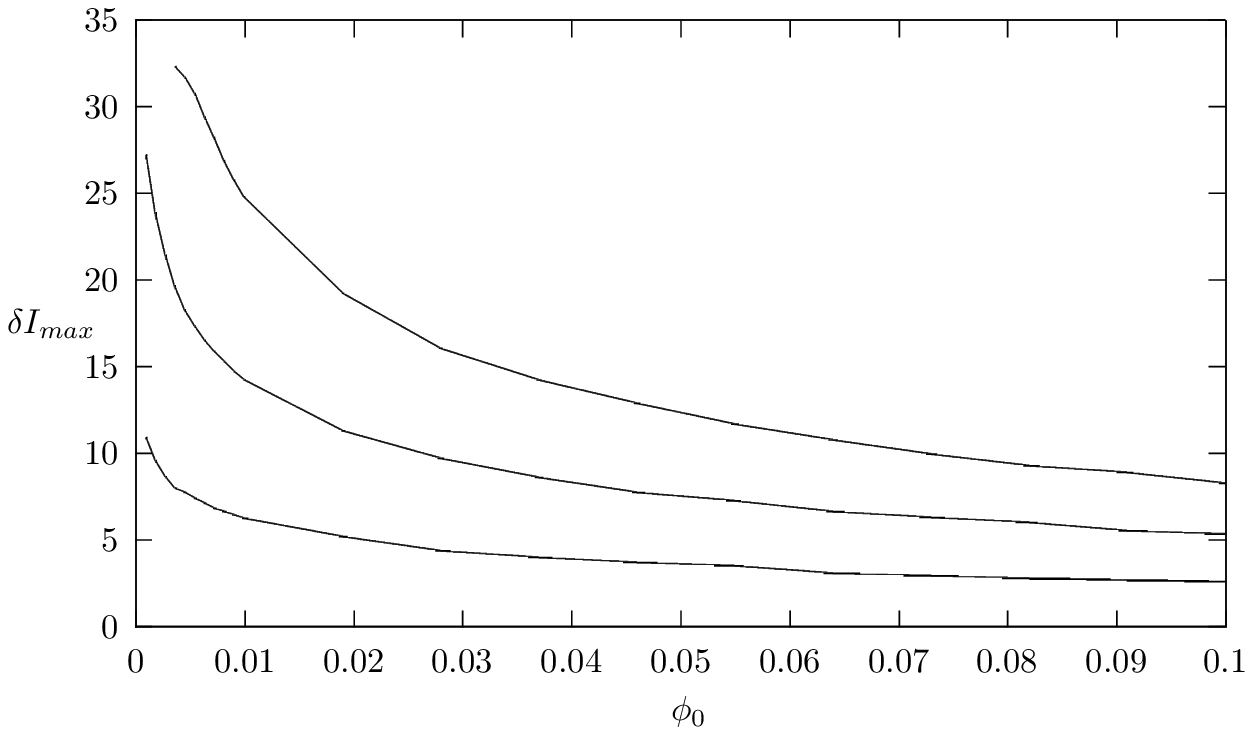}\\
c)\\
\includegraphics[scale=0.7]{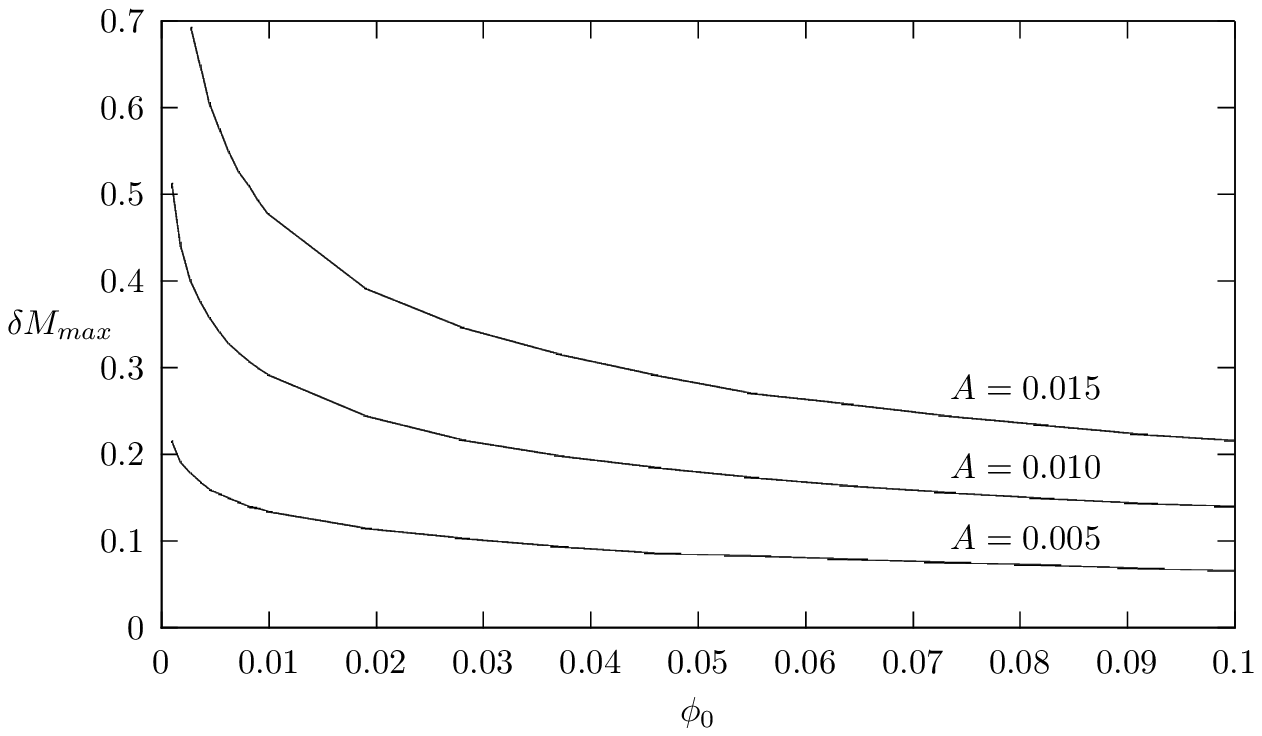} \includegraphics[scale=0.7]{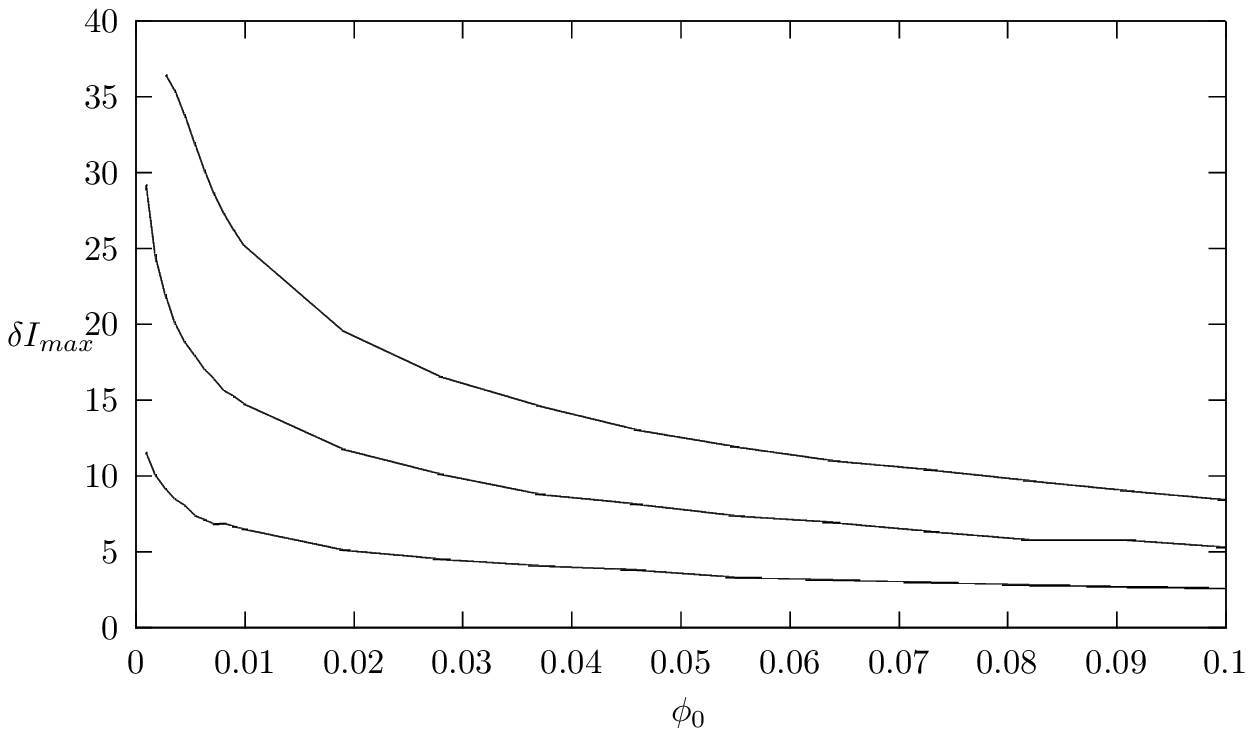}\\
\caption{The deviation of maximal star mass (left panel) and
moment of inertia (right panel) from General Relativity as
function of $\phi(0)=\phi_{0}$ for some values of parameter $A$
for a) neutron stars using SLy4 EoS, b) quark stars (EoS
(\ref{QEOS}) with $\alpha=0.28$, $B=B_{0}$), c) quark stars (EoS
(\ref{QEOS}) with $\alpha=1/3$, $B=B_{0}$).}
\end{figure}
 
\begin{table}
\caption{The relative deviations of the maximum mass and the
corresponding moment of inertia for neutron stars (SLy4 EoS). The
results for moment inertia are given in units $G^2
M^{3}_{\odot}/c^{4}=4.326\times 10^{43}$ g$\cdot$cm$^{2}$. }
    \begin{tabular}{|c||cccc|cccc|}
        \hline
\multicolumn{6}{l}{\hskip 1cm A=0.005} & \multicolumn{3}{l}{A=0.01}  \\
        \hline
        $\phi(0)$& $M_{\rm max}$ & $\Delta M_{\rm max}$[\%]& $I_{\rm max}$ &$\Delta I_{\rm max}$[\%] & $M_{\rm max}$ & $\Delta M_{\rm max}$[\%]& $I_{\rm max}$ &$\Delta I_{\rm max}$[\%] \\
        \hline
        GR & 2.05 & 0.0 & 46.2 & 0.0 & 2.05 & 0.0 & 46.2 & 0.0\\
        \hline
        0.1 & 2.08 & 1.4 &  46.9 & 1.5 & 2.11 & 2.9 &  47.8 & 3.5\\
        0.05 & 2.09 & 2.0 &  47.4 & 2.6 & 2.13 & 3.9 &  49.0 & 6.1\\
        $0.01$ & 2.11 & 2.9 &  48.8 & 5.6 & 2.19 & 6.8 &   52.5 & 13.0\\
        \hline
    \end{tabular}
\end{table}
 
\begin{table}\label{Tbl:Deviations}\caption{The relative deviations of the maximum mass and the corresponding moment of inertia for quark stars.}
    \begin{tabular}{|c||cccc|cccc|}
        \multicolumn{6}{c}{$\alpha=0.28$} & \multicolumn{3}{c}{$\alpha=1/3$} \\
        \hline
        \multicolumn{9}{l}{A=0.005}  \\
        \hline
        $\phi(0)$& $M_{\rm max}$ & $\Delta M_{\rm max}$[\%]& $I_{\rm max}$ &$\Delta I_{\rm max}$[\%] & $M_{\rm max}$ & $\Delta M_{\rm max}$[\%]& $I_{\rm max}$ &$\Delta I_{\rm max}$[\%]  \\
        \hline
        GR & 1.77 & 0.0 & 41.3 & 0.0 & 1.97 & 0.0 & 52.3 & 0.0 \\
        \hline
        0.1 & 1.84 & 4.0 &  43.8 & 6.0 & 2.04 & 3.5 &  54.8 & 4.8 \\
        0.05 & 1.86 & 5.0 &  44.8 & 8.5 & 2.06 & 4.6 &  55.9 & 6.9 \\
        $0.01$ & 1.91 & 7.9 &   47.5 & 15.0 & 2.10 & 6.6 &  58.7 & 10.9\\
        \hline
        \multicolumn{9}{l}{A=0.01}  \\
        \hline
        0.1 & 1.91 & 7.9 &  46.6 & 12.8 & 2.11 & 7.1 &  57.6 & 10.1  \\
        0.05 & 1.97 & 11.3 &  48.9 & 18.4 & 2.15 & 9.1 &  60.0 & 14.7 \\
        $0.01$ & 2.07 & 16.9 &   55.5 & 34.4 & 2.26 & 14.7 &  66.9 & 27.9 \\
        \hline
    \end{tabular}
\end{table}
 
\begin{figure}
a)\\
 \includegraphics[scale=0.7]{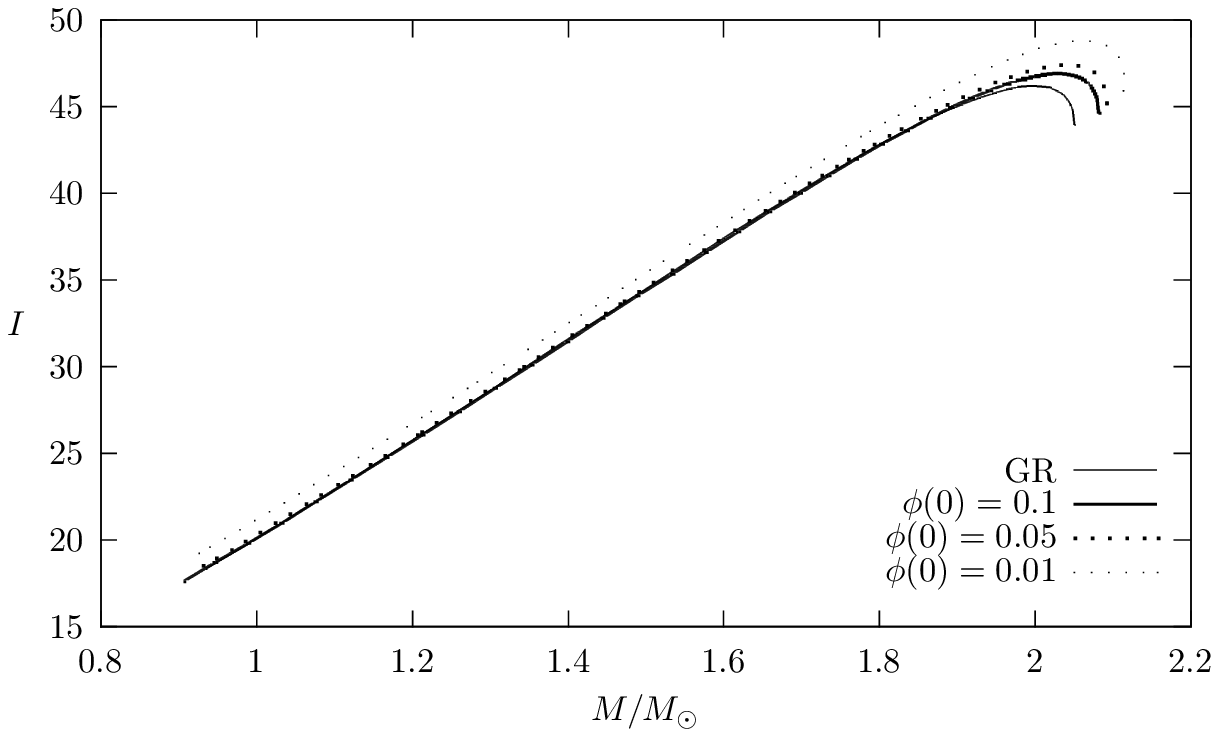} \includegraphics[scale=0.7]{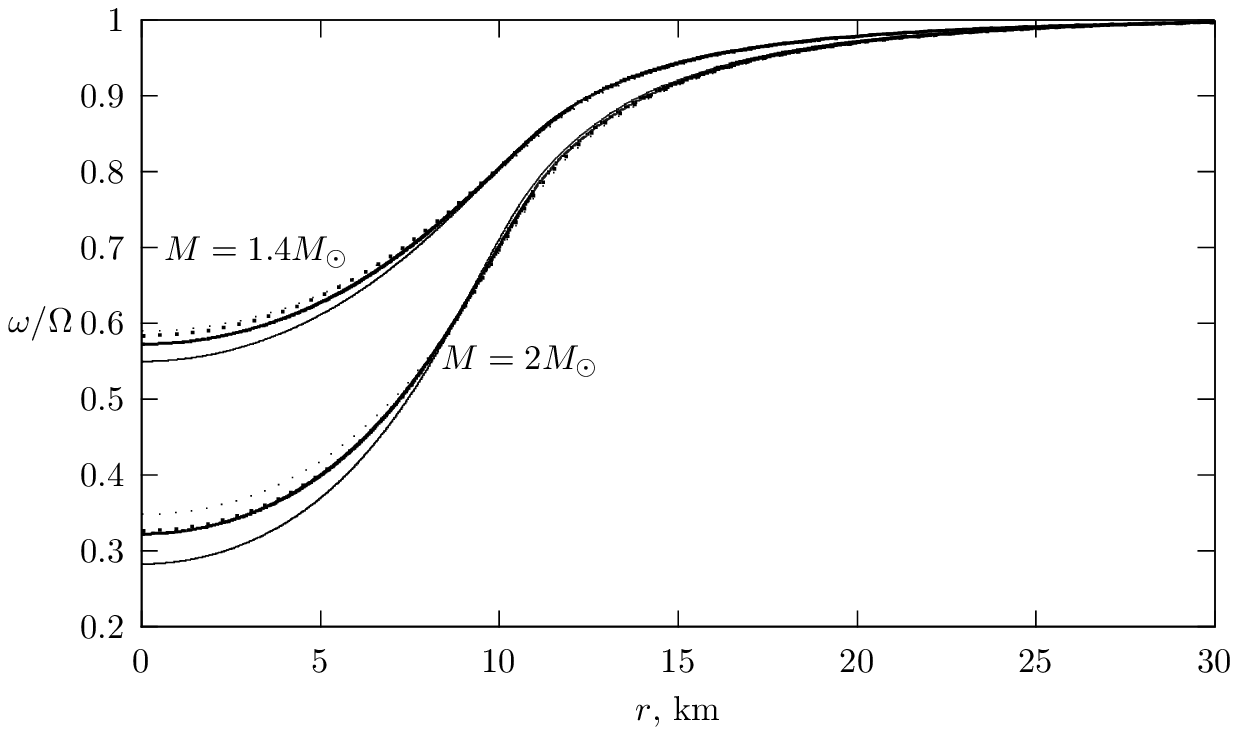}\\
b)\\
  \includegraphics[scale=0.7]{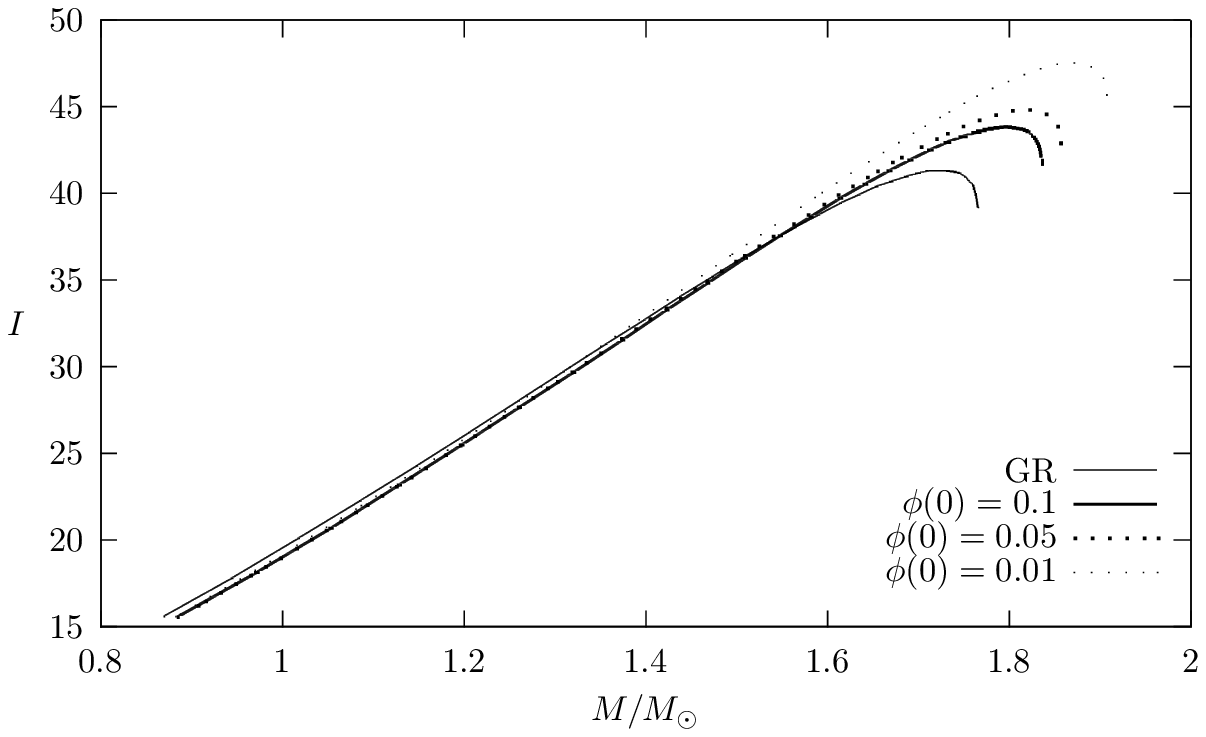} \includegraphics[scale=0.7]{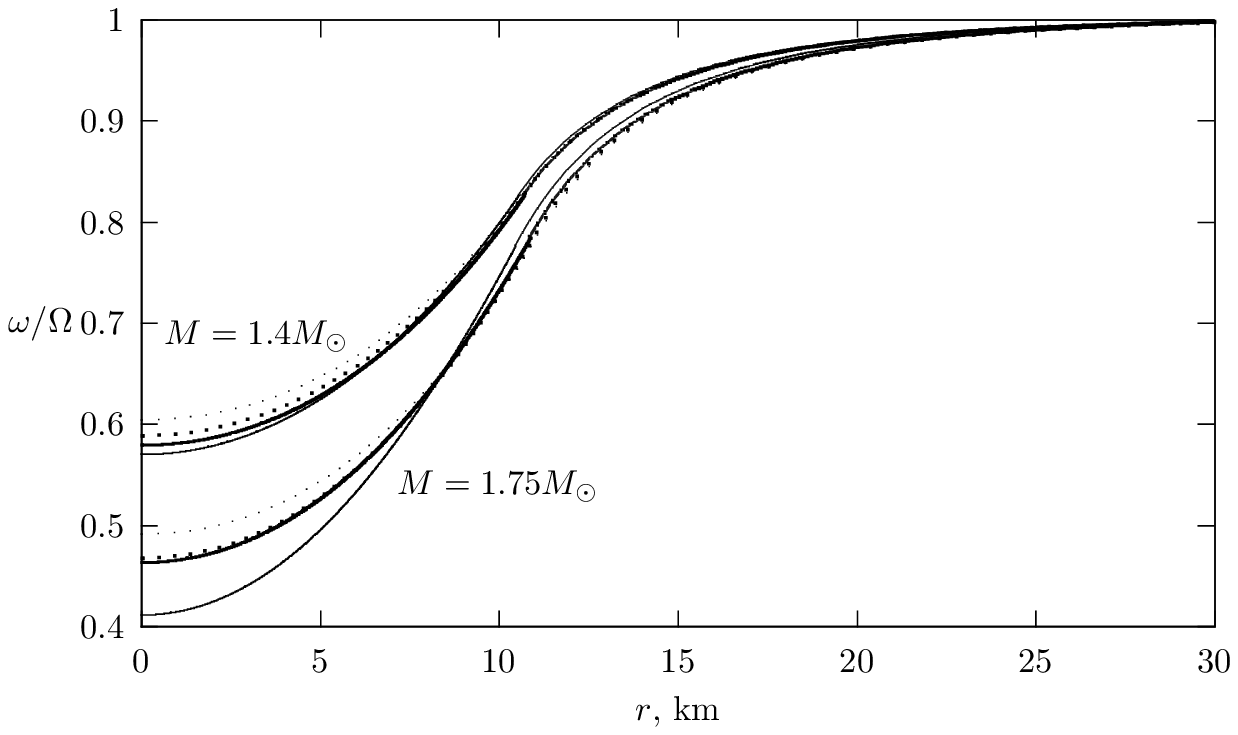}\\
c)\\
  \includegraphics[scale=0.7]{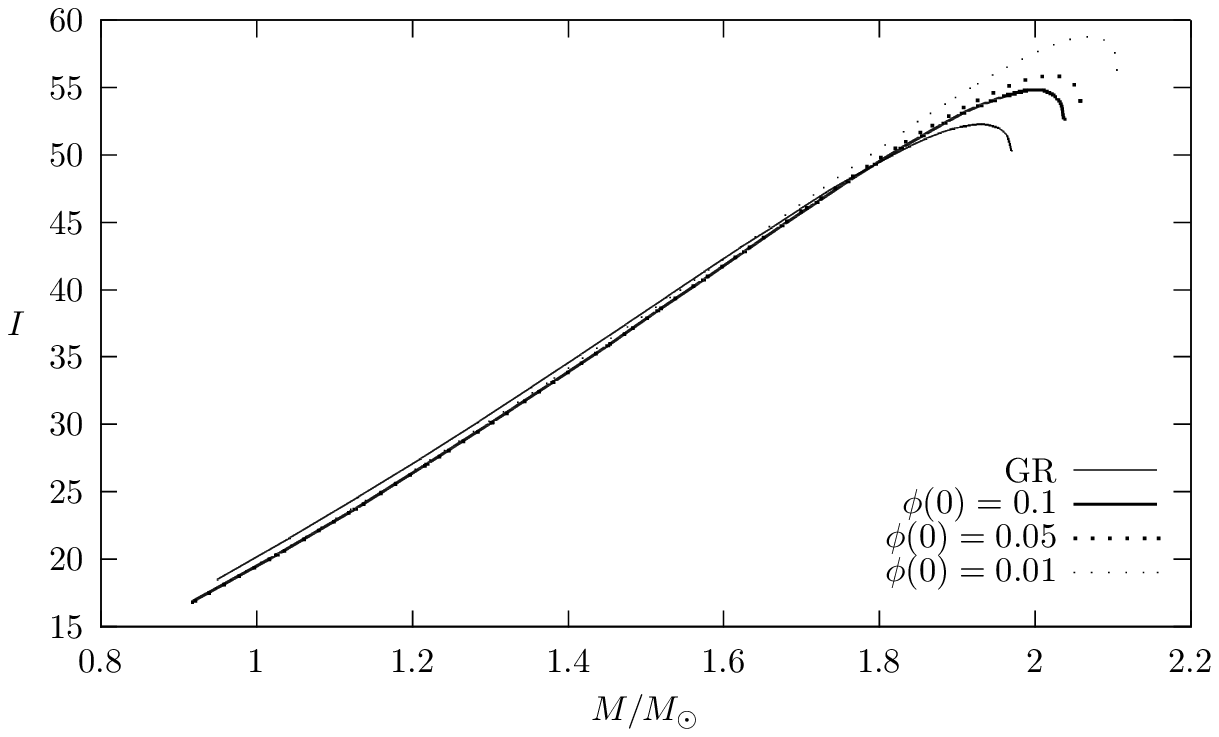} \includegraphics[scale=0.7]{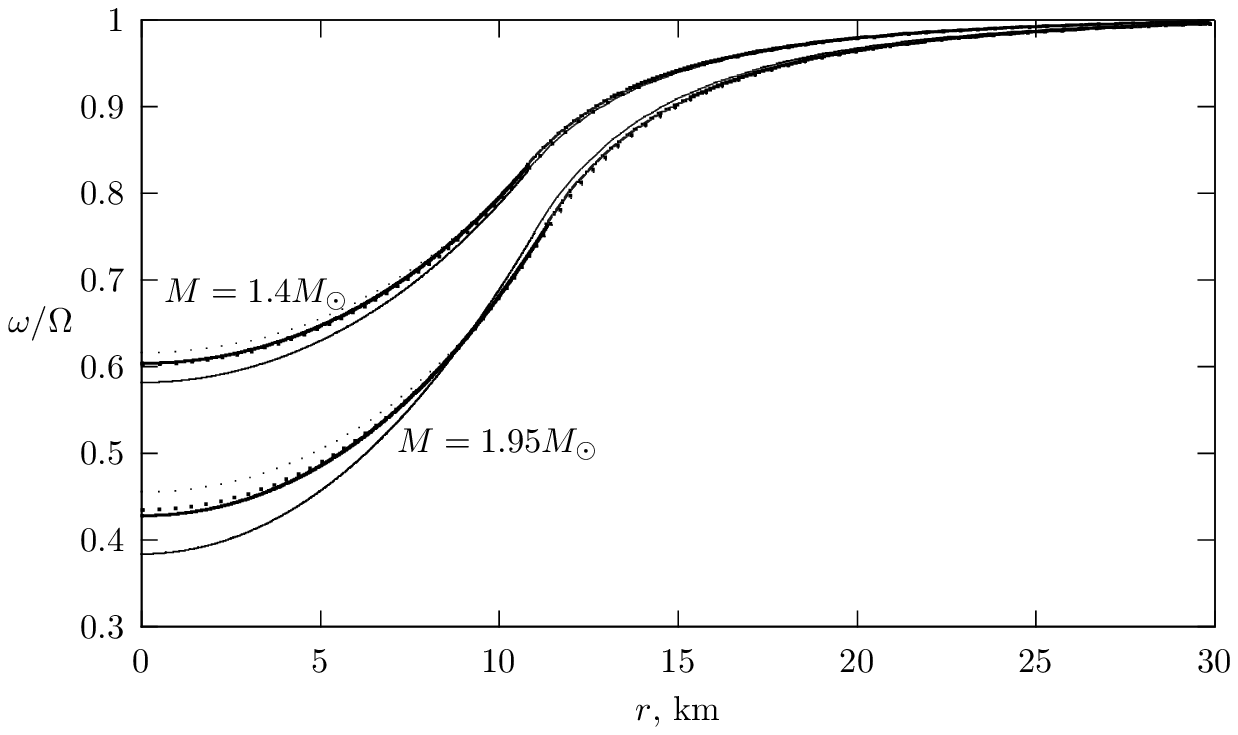}\\
  \caption{Left panel: the dependence of quark star moment of inertia from mass in mimetic GR with $V(\phi)=A\phi^{-2}$ ($A=0.005$) in comparison with General Relativity
  for a) neutron stars using SLy4 EoS, b) quark stars (EoS (\ref{QEOS}) with $\alpha=0.28$, $B=B_{0}$), c) quark stars (EoS (\ref{QEOS}) with $\alpha=1/3$, $B=B_{0}$). Right panel: the normalized metric function $\omega(r)/\Omega$ as a function of
the radial coordinate for stars with masses $M=1.4M_\odot$ and
close to maximal for corresponding EoS.}
\end{figure}
 
One may note that the relative increase of the maximum moment of
inertia is larger than the increase of the maximal mass and may
exceed the EoS uncertainty. Similar  trend has been found for
neutron star models in  scalar tensor theory of gravity and in
$R$-squared gravity \cite{Staykov2014, Yazadjiev2014}.
Determination of the moment of inertia in double pulsars as
assumed can reach high accuracy ($\sim 10$ \%) see
\cite{Kramer2009}. Therefore these observations can help us to
distinguish between General Relativity and its mimetic alternative
or set constraints on mimetic potential in the framework of
mimetic gravity.
 
\subsection{Mimetic $f(R)$ theory}
 
One should say several words about mimetic $f(R)$ ($f(R)\neq R)$
gravity. In the mathematically  equivalent scalar-tensor theory
the potential $V(\varphi)$ can be written in explicit form only
for simple $f(R)$ models. For $R$-squared gravity $f(R)=R+16\pi a
R^2$ one can obtain that
\be
V(\varphi)=\frac{1}{4a}\left(1-e^{-2\varphi/\sqrt{3}}\right)^2.
\ee
The compact star models in non-perturbative $R^{2}$ gravity have
been considered in \cite{Yazadjiev2014, Staykov2014,Capo,
Astashenok-5}. For existence of stable stars one needs the
fine-tuning for scalar curvature in the center of star. Only for
unique value of scalar curvature in center the solution of TOV
equations has required asymptotic at $r\rightarrow\infty$. In
terms of scalar-tensor description we have the fine-tuning for
scalar field $\varphi$. The gravitational mass increases with
increasing $a$. For masses $M<\sim 1.2M_{\odot}$ (Sly4 EoS) the
radii of star configurations decreases with increasing $a$.
Considering mimetic theory with scalar potential one can obtain
that this decrease can be compensated by contribution of  scalar
field. Otherwise the increase of maximal mass due to scalar field
becomes much stronger in $R^2$ gravity. As is shown in
\cite{Yazadjiev2014} the relative increase of maximal stellar mass
is only $10$ \% for $R^2$ gravity (Einstein frame description). In
$F(R)$ frame description it maybe bigger, see \cite{Capo}.
Contribution of scalar field in mimetic gravity can lead to
possible existence of extreme neutron stars with large masses.
 
For quark stars the deviation of  mass-radius relation from
General Relativity in quadratic gravity is similar to such in
mimetic model considered above. Therefore the deviations from
General Relativity in vicinity of large masses for $R^2$ mimetic
gravity can be enhanced with increasing $a$ and $V(\phi)$ in a
case of quark stars. This conclusion holds also for inertial
characteristic of compact stars.
 
\section{Conclusion}
 
In the present paper we considered realistic neutron and quark
stars in simple mimetic gravity with mimetic scalar potential. We
obtained the mass-radius relations and examined the dependence of
inertial characteristics from stellar mass.
 
For simple potentials of the form  $V(\phi)=A\phi^{-2}$ the
mass-radius relation for compact stars can considerably deviate
from the mass-radius relation in General Relativity. For neutron
stars this deviation occurs for stellar configurations with any
mass whereas for quark stars the mass-radius relation deviates
only for large masses. The deviation from GR depends on the value
of mimetic scalar in the center of star. For values of $\phi(0)$
smaller than a specific critical value $\phi_{crit}$ there exist
no stable stellar configurations. The parameter $\phi_{crit}$
depends on EoS and the form of potential. Due to the contribution
of mimetic scalar  the
 maximum  mass and the corresponding moment of inertia may increase.
This increase is considerably larger for quark stars in comparison
with the neutron stars. It should be noted that the presence of
mimetic scalar offers the possibility for the existence of stars
with low central densities $\rho<10^{15}$ g/cm$^3$ but large
masses $M>M_{\odot}$.
 
In mimetic gravity there exists a free parameter (the value of
mimetic scalar  in the center of the star as initial condition).
This freedom leads to ambiguity of mass-radius relation for given
equation of state. This ambiguity can potentially explain some
contradictions between observations and theoretical modelling of
compact stars in General Relativity. The relative deviation of the
maximal moment of inertia is approximately two times  larger than
the relative deviation of maximal stellar mass. Even for
negligible increase of mass lying within equation of state
uncertainty the increase of moment of inertia (if measured) can
help to discriminate between GR and mimetic gravity. Eventually,
the future observations of moment of inertia of compact stars will
set constraints on the models of mimetic gravity and/or convenient
modified gravity.
 
In case of $f(R)=R+aR^2$ mimetic gravity one can expect that the
increase of the maximum mass and of the maximum value of the
moment of inertia due to the presence of the scalar may become
more significant with growth of  $a$ in comparison with ordinary
$R^{2}$ gravity.

\acknowledgments
 
This work is supported in part by project 14-02-31100 (RFBR,
Russia) and 2058/60 (MES, Russia) (AVA) and  in part by MINECO
(Spain), project FIS2013-44881 and I-LINK 1019(SDO). We are grateful to Kostas Kokkotas and Salvatore Capozziello for helpful discussions and useful comments.

\bibliography{mimetic}
 
\end{document}